\documentclass[11pt]{article}
\usepackage[utf8x]{inputenc}
\usepackage{amssymb,amsmath,mathrsfs,enumerate,mathtools}
\usepackage{graphicx,rotate,multicol,braket, multirow}
\usepackage{subcaption}
\usepackage{cite}
\usepackage{braket}
\usepackage{float}
\usepackage{slashed}
\usepackage[normalem]{ulem}
\usepackage{wrapfig}
\usepackage[textfont=it,labelfont=bf]{caption}
\usepackage{bm}
\usepackage[colorlinks=true,
linkcolor=red,
urlcolor=blue,
citecolor=blue]{hyperref}
\usepackage{lineno}
\usepackage{color}
\usepackage{bbold}
\long\def\rpl#1!!#2!!{\textcolor{blue}{#1} \textcolor{blue}{#2}}

\def \order(#1){{\mathcal O} \left(#1 \right)}

\def\bar{\overline}
\def\cu{\cal U}
\def\cv{\cal V}
\def\tilde{\widetilde}

\evensidemargin=\oddsidemargin
\def\Eqn#1{Eq.\ (\ref{#1})}
\def\Eqs#1#2{Eqs.\ (\ref{#1}) and (\ref{#2})}
\def\Sect#1{Sec.\,\ref{#1}}
\allowdisplaybreaks

\title{\Large\bf 
Flavor puzzle in three Higgs-doublet models: \\Insights from BGL and lessons from flavor data
}

\author{
	\sf 
	Dipankar Das$^{a,}$\footnote{d.das@iiti.ac.in},
	Miguel Levy$^{b,}$\footnote{miguelpissarra.levy@unibas.ch},
	Anugrah M. Prasad$^{a,}$\footnote{anugrahmprasad@gmail.com}
	\\[3mm]
	\small\em
	$^a$Indian Institute of Technology (Indore), Khandwa Road, Simrol,
	Indore 453 552, India \\
	\small\em
	$^b$ Departement Physik, Universität Basel, Klingelbergstrasse 82, CH-4056 Basel, Switzerland \\ 
	\small\em
}

\date{}
\begin{document}

{\let\newpage\relax\maketitle}	

%
\renewcommand*{\thefootnote}{\arabic{footnote}}
\setcounter{footnote}{0} 


\begin{abstract}
We study a variant of the 3HDM, referred to as the BGL-3HDM,  incorporating a $U(1)_1\times U(1)_2$ symmetry, which can distinguish the primary sources of mass for different fermion generations.
In the version considered here, the Yukawa matrices in the down-quark and charged lepton sectors are diagonal, thereby eliminating tree-level FCNCs in these sectors.
FCNC interactions mediated by neutral nonstandard Higgses are confined to the up-quark sector only.
No new BSM parameters are introduced by the Yukawa sector of the model, making it as economical as the NFC versions of 3HDM with a $U(1)_1\times U(1)_2$ symmetry in terms of the number of free parameters.
However, even in the down-quark and in the charged lepton sectors, flavor diagonal but nonuniversal Higgs couplings set this model apart from the NFC versions of the 3HDM.
\end{abstract}
%

\section{Introduction}
Three Higgs-doublet models~(3HDMs) have been known to be quite useful in the context of flavor model building,
especially when used in conjunction with discrete symmetries. Such constructions are generally
accompanied by flavor changing neutral currents~(FCNCs) mediated by nonstandard scalars. Therefore, one
would usually require such nonstandard scalars to be much heavier than the electroweak scale to mitigate
the constraints from flavor data. To circumvent this situation, 3HDMs with natural flavor conservation~(NFC)~\cite{Glashow:1976nt}
have seen a recent surge in popularity~\cite{Boto:2021qgu,Chakraborti:2021bpy,Boto:2023nyi}.

In this article, instead of shunning the FCNCs altogether, we contemplate the possibility to disentangle
the sources of masses for the different generations of fermions. Quite evidently,
generation wise strong mass hierarchies in the fermion sector can then be deferred to some extent to hierarchies
among the vacuum expectation values~(VEVs) of the three doublets. Admittedly,  many efforts have already been
made along these directions~\cite{Alves:2020brq,Altmannshofer:2025pjj}, but the resulting models still contain FCNCs controlled by
new unknown parameters arising from the Yukawa sector. While the constraints arising from the FCNC couplings in the up-quark sector
can be slightly forgiving, we definitely need to be more circumspect about the FCNCs in the down-quark sector.
On this note, there exist variants of 3HDMs known as the Branco-Grimus-Lavoura~(BGL)-type 3HDM~(BGL-3HDM),
where the FCNC interactions are limited to a particular sector of 
fermions~\cite{Botella:2009pq,Serodio:2013gka,Emmanuel-Costa:2017bti,Das:2021oik}.
In the present work, we study the constructional details of such a BGL-3HDM which accommodates the following
list of features:

\begin{enumerate}[(i)]
	\item The textures of the Yukawa matrices should
	make it apparent that the charged fermions of a particular generation
	primarily receive their masses from a particular Higgs-doublet. As a result, the strong mass
	hierarchy among the different generations of fermions  can be somewhat attributed to a strong VEV hierarchy,
	$v_1\ll v_2\ll v_3$ among the three scalar doublets.
	\item For the particular variant of the BGL-3HDM considered here, the down-quark and the charged-lepton sectors
	do not contain FCNC couplings mediated by neutral scalars. Although the neutral Higgs couplings
	in these sectors are flavor diagonal, they are flavor nonuniversal. This aspect of the model
	may be exploited to selectively enhance or suppress the nonstandard Higgs couplings with
	fermions of a desired flavor.
	\item All the FCNCs, in the up-type BGL-3HDM, occur exclusively in the up-quark sector,
	with their effects being primarily controlled by the elements of the CKM matrix. This is the feature
	that carries reminiscence of the 
	BGL-type constructions~\cite{Branco:1996bq,Alves:2017xmk}.
	\item Because of the above two features, the constraints arising from the flavor data are
	quite relaxed to the extent that nonstandard particles with masses below the TeV
	scale can successfully pass through the flavor constraints if we do not insist on a very
	strong hierarchy, $v_1\ll v_2\ll v_3$, among the VEVs. 
	\item The model is, at the very least, as economical as the NFC versions of 3HDMs in
	terms of the number of physical parameters. It is quite intriguing to note that, this
	model, despite having a relatively more involved Yukawa sector, does not introduce
	any new parameters in the Yukawa sector compared to what we already have for 3HDMs
	with NFC. Therefore, when compared with the NFC avatars of 3HDMs, the predictive
	abilities are not being compromised.
\end{enumerate}
All of these features
can be incorporated within a 3HDM framework when a $U(1)\times U(1)$ symmetry is applied in a
`horizontal' manner.\footnote{It is worth noting that the original version in Ref.~\cite{Botella:2009pq}
	employed a $Z_8$ symmetry.}

Our current article is organized as follows. In \Sect{s:model} we discuss the
main features of the model. We also provide the necessary details about the
scalar and Yukawa sectors of the model. In \Sect{s:flavor} we analyze the
constraints arising from the flavor data. Our main findings are summarized
in \Sect{s:summary}.


\section{The BGL-3HDM}
\label{s:model}
The model is based on a 3HDM structure, endowed with a global $U(1)_1 \times U(1)_2$ horizontal symmetry, under which the fields transform 
as
\begin{subequations}
\label{e:transfs}
\begin{eqnarray}
	\!\!\!\!\!\!\!\!\!\!\!\!& U(1)_1 :& (Q_L)_1 \rightarrow e^{-i \psi_1}(Q_L)_1 \, , \quad (n_R)_1 \rightarrow e^{-2i \psi_1}(n_R)_1 \, , \quad  \phi_1 \rightarrow e^{i\psi_1}\phi_1 \, , \\
	\!\!\!\!\!\!\!\!\!\!\!\!&U(1)_2 :& (Q_L)_2 \rightarrow e^{-i \psi_2}(Q_L)_2 \, , \quad (n_R)_2 \rightarrow e^{-2i \psi_2}(n_R)_2  \, , \quad \phi_2 \rightarrow e^{i \psi_2}\phi_2 \, ,
\end{eqnarray}
\end{subequations}
where, $Q_L=\left(p_L~n_L\right)^T$ are the usual SM quark $SU(2)_L$ doublets, $n_R$ are the down-type $SU(2)_L$ quark singlets, and the subscripts for the fermionic fields refer to the different generations. 
For completeness, we also show below the transformations of the lepton fields 
\begin{subequations}
	\label{e:transfsl}
	\begin{eqnarray}
	\!\!\!\!\!\!\!\!\!\!\!\!& U(1)_1 :& (L_L)_1 \rightarrow e^{-i \psi_1}(L_L)_1 \, , \quad (\ell_R)_1 \rightarrow e^{-2i \psi_1}(\ell_R)_1 \, ,  \\
	\!\!\!\!\!\!\!\!\!\!\!\!&U(1)_2 :& (L_L)_2 \rightarrow e^{-i \psi_2}(L_L)_2 \, ,  \quad (\ell_R)_2 \rightarrow e^{-2i \psi_2}(\ell_R)_2 \, , 
	\end{eqnarray}
\end{subequations}
where, again,  $L_L=\left(\nu_L~\ell_L\right)^T$ are the lepton $SU(2)_L$ doublets and $\ell_R$ are the $SU(2)_L$ charged lepton singlets.\footnote{We have not included right-handed neutrinos. Thus, neutrinos are massless in this version of the model just like in the SM. However, if needed, the right-handed neutrinos can be introduced, which will be singlets under both the $SU(2)_L \times U(1)_Y$ gauge symmetry as well as the $U(1)_1 \times U(1)_2$ global symmetry. As a result, small masses for the left-handed neutrinos can be easily generated via the type-I seesaw mechanism.}
All other remaining fields transform trivially under the flavor symmetry.  
Just by a quick inspection of the transformation properties of the fields, it is already obvious that the different $U(1)$ symmetries distinguish between the different generations.  
Indeed, only fermions of the first and second generations are charged under $U(1)_1$ and $U(1)_2$, respectively, justifying their labeling.  

It is  worth mentioning that the flavor symmetry employed here does not follow the usual prescriptions that are found in NFC models, which distinguish the different sectors. This makes different Higgs doublets to be responsible for the masses of different {\em types} of fermions~\cite{Boto:2021qgu,Chakraborti:2021bpy,Boto:2023nyi}.  
In contrast, as we will see shortly, all Higgs doublets here will couple to both up- and down-type quarks.

\subsection{Scalar Sector}

We first study the scalar sector of the theory.  
This will be useful in setting up the notations that are used throughout the paper. 

Taking into account the field transformations of Eq.~\eqref{e:transfs}, the scalar potential can be written as~\cite{Ivanov:2011ae, Ferreira:2008zy}:
\begin{eqnarray}
	\label{e:potential}
	V_{{U(1) \times U(1)}}&=&
	m_{11}^2 \phi_1^\dagger \phi_1 + m_{22}^2 \phi_2^\dagger \phi_2 + m_{33}^2 \phi_3^\dagger \phi_3
	 - \left( m_{12}^2 \phi_1^\dagger \phi_2 + m_{13}^2 \phi_1^\dagger \phi_3 + m_{23}^2 \phi_2^\dagger \phi_3 +\rm{h.c.} \right)  \nonumber\\
	&&  + \lambda_1(\phi_1^\dagger \phi_1)^2 +\lambda_2(\phi_2^\dagger \phi_2)^2 + \lambda_3(\phi_3^\dagger \phi_3)^2+  \lambda_4(\phi_1^\dagger \phi_1)(\phi_2^\dagger \phi_2)  
	+ \lambda_5(\phi_1^\dagger \phi_1)(\phi_3^\dagger \phi_3) \nonumber \\
	&& + \lambda_6(\phi_2^\dagger \phi_2)(\phi_3^\dagger \phi_3) + \lambda_7(\phi_1^\dagger \phi_2)(\phi_2^\dagger \phi_1) 
	+ \lambda_8(\phi_1^\dagger \phi_3)(\phi_3^\dagger \phi_1) +\lambda_9(\phi_2^\dagger \phi_3)(\phi_3^\dagger \phi_2) \, ,
\end{eqnarray}
where $m_{12}^2$, $m_{13}^2$, and $m_{23}^2$ break the global $U(1)_1\times U(1)_2$ symmetry softly. The soft-breaking parameters are required to prevent the occurrence of massless pseudoscalars as well as to ensure safe decoupling of the nonstandard scalars~\cite{Bhattacharyya:2014oka,Faro:2020qyp}. We also assume that all the potential parameters (including VEVs) are real. 

After the electroweak symmetry breaking, it is convenient to express the scalar doublets in terms of component fields as
\begin{eqnarray}
	\phi_k = \begin{pmatrix}
		w_k^+ \\ \frac{1}{\sqrt{2}}\left(v_k+h_k+iz_k\right)
	\end{pmatrix} \, ,  ~~(k=1,2,3)
\end{eqnarray}
where, $\braket{\phi_k}=v_k/\sqrt{2}$ denotes the VEV of the $k$-th doublet. The individual VEVs are further parameterized as follows:
\begin{eqnarray}
	v_1=v\cos\beta_1\cos\beta_2 \, , \qquad v_2=v\sin\beta_1\cos\beta_2  \, , \qquad v_3=v\sin\beta_2 \, ,
\end{eqnarray}
with $v=\sqrt{v_1^2+v_2^2+v_3^2} =246$ GeV being the total electroweak VEV. The detailed analysis of the scalar potential containing the minimization of the potential and diagonalization of the mass matrices can be found in Appendix~\ref{app1:ScalarSector}.


For later convenience, we choose to work in a Higgs-basis~\cite{Georgi:1978ri, Lavoura:1994fv, Davidson:2005cw}, obtained by rotating the fields in the Lagrangian basis as follows:
\begin{eqnarray}
	\label{e:gaugetointermediate}
	\begin{pmatrix}
		H_0\\R_1\\R_2
	\end{pmatrix}= O_\beta \begin{pmatrix}
	h_1\\h_2\\h_3
\end{pmatrix} \, , \qquad \quad
	\begin{pmatrix}
		G^0 \\ A_1^\prime \\ A_2^\prime
	\end{pmatrix} =  O_\beta \begin{pmatrix}
	z_1 \\ z_2 \\ z_3
\end{pmatrix} \, , \qquad \quad
	\begin{pmatrix}
		G^\pm \\ H_1^{\prime \pm} \\H_2^{\prime \pm}
	\end{pmatrix}= O_\beta \begin{pmatrix}
	w_1^\pm \\ w_2^\pm \\ w_3^\pm
\end{pmatrix} \, ,
\end{eqnarray}
with
\begin{eqnarray}
	\label{e:Ob}
		O_\beta&=&\begin{pmatrix}
		\cos\beta_2\cos\beta_1&&\cos\beta_2\sin\beta_1&&\sin\beta_2\\-\sin\beta_1&&\cos\beta_1&&0\\
		-\cos\beta_1\sin\beta_2&&-\sin\beta_1\sin\beta_2&&\cos\beta_2	
	\end{pmatrix} \, .
\end{eqnarray}
In \Eqn{e:gaugetointermediate}, $G^0$ and $G^\pm$ stand for the neutral and charged Goldstone bosons respectively. The rest of the fields are not the physical mass eigenstates, in general. However, it should be noted that the state $H_0$ will have
SM-like couplings with the SM particles at the tree-level. The physical eigenstates
in the mass basis can be obtained via
\begin{eqnarray}
\begin{pmatrix}
G^\pm \\
H_1^\pm \\
H_2^\pm 
\end{pmatrix} 
=
O_{\gamma_1} 
\begin{pmatrix}
G^\pm \\
H_1^{\prime \pm} \\
H_2^{\prime \pm} 
\end{pmatrix}  \, , 
\qquad
\begin{pmatrix}
G^0 \\
A_1 \\
A_2 
\end{pmatrix} 
=
O_{\gamma_2} 
\begin{pmatrix}
G^0 \\
A'_1 \\
A'_2 
\end{pmatrix}  \, , 
\qquad 
\begin{pmatrix}
h \\
H_1 \\
H_2 
\end{pmatrix} 
=
O_{\alpha} O_\beta^T
\begin{pmatrix}
H_0 \\
R_1 \\
R_2 
\end{pmatrix}  \, , 
\end{eqnarray}
with $O_{\gamma_1}$,  $O_{\gamma_2}$, and $O_{\alpha}$ defined in Eqs.~\eqref{e:Ogamma1},~\eqref{e:Ogamma2}, and~\eqref{e:Oa}.  
Further details can be found in Appendix~\ref{app1:ScalarSector}.


%
For the CP-even sector, the field $H_0$ coincides with the physical mass eigenstate, $h$, with mass $m_h$ in the \emph{alignment limit}~\cite{Das:2019yad} specified by the relations,
\begin{eqnarray}
	\label{e:aligncond}
	\alpha_1=\beta_1 ~~ \rm{and} ~~ \alpha_2=\beta_2 \, .
\end{eqnarray}
Thus, in this limit, the physical eigenstate, $h$, mimics the SM Higgs boson in its tree level couplings with the SM particles. In what follows, we will work in the alignment limit so that the constraints from the Higgs signal-strengths~\cite{CMS:2022dwd} are automatically satisfied. However, it should be noted that even in the alignment limit the angle $\alpha_3$ in 
$O_\alpha$ is still a free parameter.  The main inspiration behind the Higgs basis in \Eqn{e:gaugetointermediate} is that the Yukawa couplings have relatively simpler forms in this basis as we will describe in the next subsection.


\subsection{Yukawa Sector}

Following the standard conventions, we write the Yukawa Lagrangian as
\begin{eqnarray}
	\mathscr{L}_{\rm{Y}}=-\sum_{k=1}^{3}\left[\bar{Q}_L \Gamma_k\phi_k n_R+\bar{Q}_L\Delta_k\tilde{\phi}_k p_R + \bar{L}_L \Xi_k\phi_k \ell_R\right] + \mathrm{h.c.}\, ,
\end{eqnarray}
where, $p_R$ are the up-type quark $SU(2)_L$ singlets and we have suppressed the flavor indices ($\Gamma_k$, $\Delta_k$ and $\Xi_k$ are $3\times3$ matrices in the flavor space). 
Also, $\tilde{\phi}_k = i \tau_2 \phi_k^*$, as usual, with $\tau_2$ being the second Pauli matrix.
The mass matrices for the down-type quarks, up-type quarks and charged leptons are then
given by
%
	\begin{eqnarray}
	\label{e:MnMp}
	M_n=\frac{1}{\sqrt{2}}\sum_{k=1}^{3}v_k \, \Gamma_k \, ,  \qquad M_p=\frac{1}{\sqrt{2}}\sum_{k=1}^{3}v_k \, \Delta_k \, ,  \qquad
	M_\ell=\frac{1}{\sqrt{2}}\sum_{k=1}^{3}v_k \, \Xi_k \, .
	\end{eqnarray}

The textures of $\Gamma$, $\Delta$ and $\Xi$ matrices that follow from the transformations of \Eqs{e:transfs}{e:transfsl} are given below:
\begin{subequations}
	\begin{eqnarray}
		\label{e:gamtexture}
		&&
			\Gamma_1=\begin{pmatrix}
				y_1^d & 0 & 0 \\ 0 & 0 & 0 \\ 0 & 0 & 0
			\end{pmatrix} , \quad
			 	\Gamma_2=\begin{pmatrix}
				0 & 0 & 0 \\ 0 & y_2^d & 0 \\ 0 & 0 & 0
			\end{pmatrix} , \quad
			 	\Gamma_3=\begin{pmatrix}
				0 & 0 & 0 \\ 0 & 0 & 0 \\ 0 & 0 & y_3^d
			\end{pmatrix} ,
\\
		\label{e:deltexture}
		&&
			\Delta_1=\begin{pmatrix}
				a_1 & b_1 & c_1 \\ 0 & 0 & 0 \\ 0 & 0 & 0
			\end{pmatrix} , \quad
			 	\Delta_2=\begin{pmatrix}
				0 & 0 & 0 \\ a_2 & b_2 & c_2 \\ 0 & 0 & 0
			\end{pmatrix}  , \quad
			 	\Delta_3=\begin{pmatrix}
				0 & 0 & 0 \\ 0 & 0 & 0 \\ a_3 & b_3 & c_3
			\end{pmatrix} , \\
		\label{e:xitexture}
&&
\Xi_1=\begin{pmatrix}
y_1^\ell & 0 & 0 \\ 0 & 0 & 0 \\ 0 & 0 & 0
\end{pmatrix}  , \quad
 	\Xi_2=\begin{pmatrix}
0 & 0 & 0 \\ 0 & y_2^\ell & 0 \\ 0 & 0 & 0
\end{pmatrix} , \quad
 	\Xi_3=\begin{pmatrix}
0 & 0 & 0 \\ 0 & 0 & 0 \\ 0 & 0 & y_3^\ell
\end{pmatrix}  ,
\end{eqnarray}
\end{subequations}
An intriguing feature already emerges from these textures. Since 
$\Gamma_k$, $\Xi_k$, and consequently $M_n$ and $M_\ell$
are already diagonal, it is evident that the masses of each generation in the down-quark and charged lepton sectors will be governed by the VEV of a dedicated scalar doublet. Furthermore, if the VEVs of the scalars are hierarchical ($v_1 \ll v_2 \ll v_3$), the hierarchical structure of the Yukawa coefficients in the down-quark and charged lepton sectors will be significantly diluted.

Although less immediately apparent than in the down-quark sector, it can be demonstrated that the same VEV hierarchy will have analogous effects in the up-quark sector. Specifically, the VEV of a particular scalar doublet will serve as the primary source of mass for each generation of quarks in the up-quark sector as well. Further details on the up-quark sector can be found in subsection~\ref{app:Diagonalization}. For now, however, we will focus on the FCNC couplings that arise.


To understand the structure of the FCNCs, let us first determine the couplings of the physical quarks to the scalar fields defined in Eq.~\eqref{e:gaugetointermediate}.  
The couplings to the physical eigenstates can then be easily derived through appropriate rotations originating solely from the scalar sector.
To begin with, let us define the physical quark fields via the following unitary rotations:
\begin{eqnarray}
	d_L={\cu}_d \, n_L \, , \quad  d_R={\cv}_d \, n_R \, , \qquad 
	u_L={\cu}_u \, p_L \, , \quad  u_R={\cv}_u \, p_R \,, \qquad
	e_L={\cu}_\ell \, \ell_L \, , \quad  e_R={\cv}_\ell \, \ell_R \,, 
\end{eqnarray}
where, $u\equiv(u~c~t)^T$, $d\equiv(d~s~b)^T$ and $e\equiv(e~\mu~\tau)^T$ denote the physical fermion fields.
After these rotations, the diagonal mass matrices will be obtained as follows:
\begin{subequations}
\label{e:DdDu}
\begin{eqnarray}
	&& D_d = {\cu}_d\cdot M_n\cdot {\cv}_d^\dagger = {\rm diag}(m_d,~m_s, ~m_b) \,,  \\
	&& D_u = {\cu}_u\cdot M_p\cdot {\cv}_u^\dagger = {\rm diag}(m_u,~m_c, ~m_t) \,,  \\
	&& D_\ell = {\cu}_\ell\cdot M_\ell\cdot {\cv}_\ell^\dagger = {\rm diag}(m_e,~m_\mu, ~m_\tau) \,,
\end{eqnarray}
\end{subequations}
and the CKM matrix will be given by $V= {\cu}_u{\cu}_d^\dagger$.
Since $M_{n,\ell}$ are already diagonal, we can take ${\cu}_{d,\ell}$ to be the identity matrices, whereas it is understood that ${\cv}_{d,\ell}$ will provide the necessary rephasings such that the elements of $D_{d,\ell}$ are real and positive.  
Thus, since we are already in a basis where the down-sector is diagonal, we may write,  
\begin{eqnarray}
	\label{e:GammaShapes}
	 \dfrac{1}{\sqrt{2}}\sum_{k=1}^{3}v_k \, \Gamma_k = \dfrac{1}{\sqrt{2}} \mathrm{diag}\left( v_1\, y^d_1 \, , \, \, v_2 \, y^d_2 \, , \, \, v_3 \,  y^d_3 \right)=\mathrm{diag}\left( m_d \, , \, \, m_s \, , \, \, m_b \right) \, . 
\end{eqnarray}
Consequently, the CKM mixing arises exclusively from the up-sector as follows:
\begin{eqnarray}
	\label{e:VCKMdef}
	V= {\cu}_u \cdot {\cu}_d^\dagger \equiv {\cu}_u \, .
\end{eqnarray}

With this insight, let us first focus on the neutral Higgs couplings in the CP-even sector. We start with the relevant part of Yukawa Lagrangian: 
\begin{eqnarray}
	-\mathscr{L}_{\mathrm{Y}}^{\mathrm{CP-even}}&=& \frac{1}{\sqrt{2}}\sum_{k=1}^{3}\bigg[\bar{n}_L \, \Gamma_k  \, h_k  \, n_R + \bar{p}_L \, \Delta_k \,  h_k  \, p_R \bigg] + \mathrm{h.c.} \nonumber \\
	&\equiv&\bar{d}_L \left(\frac{1}{\sqrt{2}}\sum_{k=1}^{3}\Gamma_k  \,  h_k \right) d_R + \bar{u}_L \, {\cu}_u \left(\frac{1}{\sqrt{2}}\sum_{k=1}^{3}\Delta_k  \, h_k\right) {\cv}_u^\dagger \, u_R + \mathrm{h.c.} 
\end{eqnarray}
Using the definitions of~\Eqn{e:gaugetointermediate}, we can decompose $h_k$ as 
\begin{eqnarray}
h_k = \left(O_\beta\right)_{1k} H_0  + \left(O_\beta\right)_{2k} R_1  + \left(O_\beta\right)_{3k} R_2 \, , 
\end{eqnarray}
so that the Yukawa Lagrangian in the CP-even sector may be rewritten as
\begin{eqnarray}
\label{e:CPevenCouplings}
	-\mathscr{L}_{\mathrm{Y}}^{\mathrm{CP-even}} & = & 
	\dfrac{H_0}{\sqrt{2}} \left[ \bar{d}_L \left( \sum_{k=1}^{3} (O_\beta)_{1k} \Gamma_k \right)  d_R +  \bar{u}_L \, {\cu}_u \left( \sum_{k=1}^{3} (O_\beta)_{1k} \Delta_i \right) {\cv}_u^\dagger \, u_R \right] \nonumber \\
	& & +\dfrac{R_1}{\sqrt{2}} \left[  \bar{d}_L  \left( \sum_{k=1}^{3} (O_\beta)_{2k} \Gamma_k \right)  d_R + \bar{u}_L \,  {\cu}_u \left( \sum_{k=1}^{3} (O_\beta)_{2k} \Delta_k \right) {\cv}_u^\dagger \, u_R \right] \nonumber \\
	& & +\dfrac{R_2}{\sqrt{2}} \left[ \bar{d}_L  \left( \sum_{k=1}^{3} (O_\beta)_{3k} \Gamma_k \right)  d_R   + \bar{u}_L \, {\cu}_u \left( \sum_{k=1}^{3} (O_\beta)_{3k} \Delta_k \right) {\cv}_u^\dagger \, u_R \right] +  \rm{h.c.}
\end{eqnarray}
By virtue of being a Higgs basis, the couplings of $H_0$ with the physical
quarks simplify to 
 \begin{eqnarray}
 	-\mathscr{L}_{\mathrm{Y}}^{H^0}&=&  \dfrac{H_0}{v} \bigg[ \bar{d}_L \, D_d \, d_R + \bar{u}_L \, D_u \, u_R \bigg] + \mathrm{h.c.} =\dfrac{H_0}{v}\bigg[ \bar{d} \, D_d \, d +\bar{u} \, D_u \,  u \bigg]\, ,
\end{eqnarray}
elucidating the fact that $H_0$ has SM-like couplings. This is a general feature
of 3HDMs and this is why the constraints from the Higgs-signal strengths are satisfied in the alignment limit.

Since the model does not guarantee the simultaneous diagonalizability of all Yukawa matrices in the up-quark sector, the nonstandard neutral scalars will exhibit tree-level FCNC interactions with the up-type quarks. In fact,  
the quark couplings of $R_1$ and $R_2$ can be easily read from \Eqn{e:CPevenCouplings} as
\begin{eqnarray}
	-\mathscr{L}_{\mathrm{Y}}^{R_i} &=& 
	R_1 \bigg[ \bar{d}_L \, N_1^d \, d_R + \bar{u}_L \, N_1^u \, u_R \bigg] + R_2 \bigg[ \bar{d}_L \, N_2^d \, d_R + \bar{u}_L \, N_2^u \, u_R \bigg] + \mathrm{h.c.}  \nonumber \\
	&=& R_1 \bigg[ \bar{d} \left(N_1^d \, P_R + {N_1^d}^\dagger \, P_L \, \right) d + \bar{u} \left(N_1^u \, P_R + {N_1^u}^\dagger \, P_L \right) u \bigg] \nonumber \\
	&+&  R_2 \bigg[ \bar{d} \left(N_2^d \, P_R + {N_2^d}^\dagger \, P_L \, \right) d + \bar{u} \left(N_2^u \, P_R + {N_2^u}^\dagger \, P_L \right) u \bigg] \,,
\end{eqnarray}
where, we have introduced the following shorthands for the matrices that govern
the neutral current interactions in the quark Yukawa sectors~\cite{Botella:2009pq}:
\begin{subequations}
\label{e:NuNddef}
\begin{eqnarray}
	\label{e:Nu1d1def}
	N_1^d=\frac{1}{\sqrt{2}}\sum_{k=1}^{3} (O_\beta)_{2k} \Gamma_k \, ,  \qquad 
	N_1^u={\cu}_u \left(\frac{1}{\sqrt{2}}\sum_{k=1}^{3} (O_\beta)_{2k} \Delta_k \right) {\cv}_u^\dagger \, ,  \\
	\label{e:Nu2d2def}
	N_2^d=\frac{1}{\sqrt{2}}\sum_{k=1}^{3} (O_\beta)_{3k} \Gamma_k \, ,  \qquad 
	N_2^u={\cu}_u \left(\frac{1}{\sqrt{2}}\sum_{k=1}^{3} (O_\beta)_{3k} \Delta_k \right) {\cv}_u^\dagger \, .  
\end{eqnarray}
\end{subequations}
Before examining the specific forms of these matrices, we note that the basis choice in \Eqn{e:gaugetointermediate} acts on the scalar doublets as a whole. Consequently, these same matrices will also govern the couplings of quarks with the CP-odd and charged scalars, albeit through different combinations.
To see this explicitly in the pseudoscalar sector, we use \Eqn{e:gaugetointermediate} to write
\begin{eqnarray}
z_k = \left(O_\beta\right)_{1k} G_0  + \left(O_\beta\right)_{2k} A'_1  + \left(O_\beta\right)_{3k} A'_2 \,.
\end{eqnarray}
Following the same steps as in the case of the CP-even sector, we can write the
pseudoscalar Yukawa couplings as follows:\footnote{The Goldstone Yukawa couplings take the same form as those in the Standard Model and are not explicitly shown here.}
\begin{eqnarray}
	-\mathscr{L}_{\mathrm{Y}}^{\mathrm{CP-odd}} &=& 
	\frac{1}{\sqrt{2}} \sum_{k=1}^{3} \bigg[ \bar{n}_L \, \Gamma_k  \, (i z_k)  \, n_R + \bar{p}_L \, \Delta_k  \, (i z_k)  \, p_R \bigg] + \mathrm{h.c.} \nonumber \\
	&=& i\, A^\prime_1 \bigg[ \bar{d} \left(N_1^d \, P_R - {N_1^d}^\dagger \, P_L \, \right) d + \bar{u} \left(N_1^u \, P_R - {N_1^u}^\dagger \, P_L \right) u \bigg] \nonumber \\
	&  +& i \, A^\prime_2 \bigg[ \bar{d} \left(N_2^d \, P_R - {N_2^d}^\dagger \, P_L \, \right) d + \bar{u} \left(N_2^u \, P_R - {N_2^u}^\dagger \, P_L \right) u \bigg] \, .
\end{eqnarray}
Finally, Yukawa interactions for the charged scalars can be extracted in a similar manner as follows:
\begin{eqnarray}
	-\mathscr{L}_{\mathrm{Y}}^\mathrm{charged} =
	 \sqrt{2} H_1^{\prime+} \left[ \bar{u}  \left( V  N_1^d P_R - {N_1^u}^\dagger V  P_L \right) d \right]  +
	 \sqrt{2} H_2^{\prime+} \left[ \bar{u}  \left( V  N_2^d P_R - {N_2^u}^\dagger V  P_L \right) d \right] + \mathrm{h.c.}
\end{eqnarray}

\subsection{Up-type quark masses in the BGL-3HDM}
\label{app:Diagonalization}
Given the textures of the Yukawa matrices, it is natural to wonder how these relate to the physical fermionic masses and mixings.  
Here, we will sketch the details of how some of the hierarchies
in the quark masses and mixings can be transferred to the scalar sector.  
For an intuitive understanding of the procedure, we have assumed all the Yukawa parameters to be real. However, we have also checked that the phase of the CKM matrix can be easily accommodated once we allow the relevant parameters to be complex.

First, we focus on the down-quark sector due to its simplicity. 
From \Eqn{e:GammaShapes}, we can readily determine that the Yukawa coefficients necessary for the down-type masses are\footnote{The Yukawa couplings ($y_k^\ell$) 
in the charged lepton sector will have similar expressions.}
\begin{eqnarray}
y_1^d = \dfrac{\sqrt{2} m_d}{v_1} \, , \qquad 
y_2^d = \dfrac{\sqrt{2} m_s}{v_2} \, , \qquad 
y_3^d = \dfrac{\sqrt{2} m_b}{v_3} \, .
\end{eqnarray}
Then, in order to reduce hierarchies among the Yukawas to the greatest extent, we must have \mbox{$v_1 \ll v_2 \ll v_3$}.  
For simplicity, define two small parameters, $\epsilon_1$ and $\epsilon_2$:\footnote{For small values of $\epsilon_{1,2}$, we will have $\tan\beta_1= 1/\epsilon_1$ and $\tan\beta_2 \approx 1/\epsilon_2$.}
\begin{eqnarray}
\label{e:eps12}
v_1 = \epsilon_1 \, \epsilon_2 \, v_3 \, , \qquad \quad v_2 = \epsilon_2 \, v_3\, .
\end{eqnarray}  

On the other hand, the up sector needs to account for both the up-type masses, as well as for the observed quark mixing.  
Using \Eqn{e:MnMp} in conjunction with the textures of $\Delta_i$ in \Eqn{e:deltexture}, we can write
\begin{eqnarray}
\label{e:videli}
M_p = \frac{1}{\sqrt{2}}
\begin{pmatrix} 
a_1 v_1 & b_1 v_1 & c_1 v_1 \\
a_2 v_2 & b_2 v_2 & c_2 v_2 \\
a_3 v_1 & b_3 v_3 & c_3 v_3 
\end{pmatrix} \, , 
\end{eqnarray}
such that the relevant Hermitian combination is given by 
\begin{eqnarray}\label{e:Hu}
 H_u = M_p M_p^\dagger & = & \dfrac{1}{2}
\begin{pmatrix} 
v_1^2 \left( a_1^2+b_1^2+c_1^2 \right) &  v_1 v_2 \left( a_1 a_2+b_1 b_2+c_1 c_2 \right)  &  v_1 v_3 \left( a_1 a_3+b_1 b_3+c_1 c_3 \right)  \nonumber \\
. & v_2^2 \left( a_2^2+b_2^2+c_2^2 \right) &  v_2 v_3 \left( a_2 a_3+b_2 b_3+c_2 c_3 \right)  \\
. & .& v_3^2 \left( a_3^2+b_3^2+c_3^2 \right)  
\end{pmatrix} \\
& \equiv &
\begin{pmatrix}
v_1^2 \, y_{11} & v_1 \, v_2 \, y_{12} & v_1 \,  v_3 \, y_{13} \\
. & v_2^2 \, y_{22} & v_2 \, v_3 \, y_{23} \\
. & . & v_3^2 \, y_{33}
\end{pmatrix} \, ,  
\end{eqnarray}
where, for simplicity, we have defined
\begin{eqnarray}\label{e:yjkdef}
	y_{jk}=\frac{1}{2}\left(a_ja_k+b_jb_k+c_jc_k\right) \,, \quad {\rm with} \quad j,k=1,2,3 \,.
\end{eqnarray}
It should be emphasized that these $y_{jk}$'s are the parameters of physical
interest. Our goal is to show that even with this slightly more involved mass
squared matrix, the VEVs $v_1$, $v_2$ and $v_3$ remain the dominant sources
of masses for $m_u$, $m_c$ and $m_t$ respectively.
First, note that for $v_1\ll v_2\ll v_3$ and non-hierarchical $y_{jk}$, the matrix in \Eqn{e:Hu} is approximately diagonal.  
In this way, we can anticipate that the rotation angles necessary to bring $H_u$ in its diagonal form are small. This also suggests that the hierarchies in the quark mixing angles can be traced back to the hierarchy of the VEVs. 
To explore this in greater detail, we will reverse-engineer the matrix $H_u$ from the observed values 
of quark masses and mixings and express its elements as powers of the Wolfenstein parameter, $\lambda\approx 0.22$. We will then compare this reverse-engineered form of $H_u$ 
with the expression derived in \Eqn{e:Hu}.

To begin with, we use \Eqs{e:DdDu}{e:VCKMdef} to write,
\begin{eqnarray}
\label{e:matching1}
H_u = V \, D_u^2 \,  V^\dagger \, , 
\end{eqnarray}
with $V$ being the CKM matrix.  
To ensure perfect unitarity of the CKM matrix, we use the standard parametrization: 
\begin{subequations}
	\label{e:stdP}
	\begin{eqnarray}
		V &=& {\cal R}_{23} \cdot  {\cal R}_{13} \cdot {\cal R}_{12} \,,
	\end{eqnarray}
with, 
\begin{equation}
	\label{e:RCKM}
	{\cal R}_{12} = \begin{pmatrix}
		c_{\theta_{12}} & s_{\theta_{12}} & 0 \\
		-s_{\theta_{12}} & c_{\theta_{12}} & 0 \\
		0 & 0 & 1 \end{pmatrix}
	\,, \quad {\cal R}_{13} = \begin{pmatrix}
		c_{\theta_{13}} & 0 & e^{-i \delta} s_{\theta_{13}}  \\
		0 & 1 & 0 \\
		-e^{i \delta} s_{\theta_{13}} & 0 & c_{\theta_{13}} 
	\end{pmatrix}\,,  \quad
	{\cal R}_{23} = \begin{pmatrix}
		1 & 0 & 0 \\
		0 & c_{\theta_{23}} &  s_{\theta_{23}}  \\
		0 & -s_{\theta_{23}} & c_{\theta_{23}}
	\end{pmatrix},
\end{equation}
\end{subequations}
where we use the usual notation. where $c_x \equiv \cos x$, $s_x  \equiv \sin x$, and $t_x \equiv  \tan x$, throughout the paper.
For an intuitive understanding, we will ignore the small CP-violating phase and therefore we will set $\delta\approx 0$.  
To make the connection with the Wolfenstein parametrization~\cite{Wolfenstein:1983yz}
apparent, one should identify~\cite{ParticleDataGroup:2024cfk}
\begin{eqnarray}\label{e:dict}
\sin\theta_{12} = \lambda \, , \qquad \sin\theta_{23} = A \lambda^2 \, , \qquad \sin\theta_{13} = A \lambda^3 \rho \, . 
\end{eqnarray}
With our objective in mind, we define
\begin{eqnarray}
\label{e:suppressions}
y_u \, \lambda^7 = \frac{\sqrt{2} \, m_u}{v}\, , \qquad  y_c \, \lambda^3 = \frac{\sqrt{2} \, m_c}{v}\, , \qquad y_t = \frac{\sqrt{2} \, m_t}{v} \, , 
\end{eqnarray}
with $\lambda$ being the Wolfenstein expansion parameter, such that $y_{u,c,t} \sim \mathcal{O}(1)$.

Finally, using \Eqs{e:dict}{e:suppressions}, we find
\begin{eqnarray}
\label{e:matching0}
 V \, D_u^2 \,  V^\dagger \,  \approx v^2
\begin{pmatrix}
A^2 \, \lambda^6 \, \rho^2 \, y_t^2 & -A^2 \, \lambda^5 \, \rho \, y_t^2 & A \, \lambda^3 \, \rho \, y_t^2\\
-A^2 \, \lambda^5 \, \rho \, y_t^2 & A^2 \, \lambda^4 \, y_t^2& -A \, \lambda^2 \, y_t^2\\
A \, \lambda^3 \, \rho \, y_t^2 & -A \, \lambda^2 \, y_t^2 & y_t^2 
\end{pmatrix} \, ,
\end{eqnarray}
where we have kept only the leading order terms in $\lambda$.

Now, combining \Eqs{e:eps12}{e:Hu} we may write, 
\begin{eqnarray}
\label{e:matching2}
H_u = v_3^2
\begin{pmatrix}
\epsilon_1^2 \epsilon_2^2 \, y_{11} & \epsilon_1 \epsilon_2^2 \, y_{12}& \epsilon_1 \epsilon_2 \, y_{13}\\
. & \epsilon_2^2 \, y_{22}& \epsilon_2 \, y_{23}\\
. & .& y_{33} 
\end{pmatrix} \, ,
\end{eqnarray}
where $v_3\approx v$ will be assumed whenever $\epsilon_1,\epsilon_2\ll1$ holds true.
In fact, looking at \Eqn{e:matching1} and comparing \Eqs{e:matching0}{e:matching2}, we can indeed infer that it is desirable to have
\begin{eqnarray}
\label{e:vevorders}
\epsilon_1 \sim \mathcal{O}(\lambda) \, , \qquad \quad \epsilon_2 \sim \mathcal{O}(\lambda^2)\,,
\end{eqnarray}
resulting in the hierarchy $v_1\ll v_2\ll v_3$.\footnote{
One should note that such a hierarchy, although desirable, may not be essential. We can easily see this in the limit $V=1$
which is not very far from reality. In this limit, \Eqn{e:matching1} reduces to simply $H_u\equiv D_u^2$. Therefore, comparison
with \Eqn{e:matching2} will lead to $y_{jk}=0$ for $j\ne k$, and $m_u^2=y_{11}\epsilon_1^2 \epsilon_2^2v_3^2\equiv y_{11}v_1^2$,
$m_c^2=y_{22} \epsilon_2^2v_3^2\equiv y_{22}v_2^2$ and $m_t^2=y_{33} v_3^2$ clearly depicting the segregation of the sources
of masses for the different generations of up-type quarks. 
}

At this point, it is important to discuss the significance  of \Eqn{e:vevorders} in the context of \Eqs{e:matching0}{e:matching2}.  
By comparing $H_u$ and $V D_u^2 V^\dagger$, we observe that even if the coefficients $y_{jk}$ are of the similar order, the quark mixing angles will approximately match the experimentally measured values due to the hierarchical structure of the VEVs.  
Therefore, the Yukawa coefficients in this model do not need to conspire among themselves to account for the smallness of the mixing angles, in contrast to the situation in the SM.  

To obtain the physical masses in the up-sector, we need to proceed perturbatively.
The top mass can be extracted by simply comparing the leading terms of the (3,3) elements in \Eqs{e:matching0}{e:matching2}:
\begin{eqnarray}
\label{e:ytLO}
y_t^2\, v^2 \approx y_{33}\, v_3^2 \,.
\end{eqnarray}
Thus, using \Eqn{e:suppressions}, we can write
\begin{eqnarray}
	\label{e:mt}
	m_t \equiv \frac{1}{\sqrt{2}}\, y^u_3 \, v_3 \,, \quad {\rm with} \quad 
	y^u_3 = \sqrt{y_{33}} \,,
\end{eqnarray}
which implies that $v_3$ will be the dominant source for the mass of top quark.

From \Eqn{e:matching0}, it is clear that all the leading contributions to $H_u$ stem from the top Yukawa. 
As such, extracting the charm (and up) Yukawas will not be  as straightforward.  
In fact, doing so requires going beyond the leading order in $\lambda$.
To extract the charm quark mass
we need to focus on the (2,2), (2,3), and (3,3) elements of $(V^\dagger D_u^2 V)$
in \Eqn{e:matching0}, and keep terms up to $\order(\lambda^6)$:
\begin{subequations}
\begin{eqnarray}
\frac{1}{v^2} \left(  V^\dagger \, D_u^2 \,  V \right)_{3,3} &=& y_t^2 - A^2 \lambda^4 y_t^2 - A^2 \lambda^6 \rho^2 y_t^2 + \mathcal{O}(\lambda^{10}) \, , \label{e:yc1}\\
\frac{1}{v^2} \left(  V^\dagger \, D_u^2 \,  V \right)_{2,2} &=& A^2 \lambda^4 y_t^2 + \lambda^6 y_c^2   + \mathcal{O}(\lambda^8) \, , \label{e:yc2}\\
\frac{1}{v^2} \left(  V^\dagger \, D_u^2 \,  V \right)_{2,3} &=& -A \lambda^2 y_t^2 + \frac{1}{2} A^3 \lambda^6 y_t^2 + \mathcal{O}(\lambda^8) \, . \label{e:yc3}
\end{eqnarray}
\end{subequations}
Inspecting \Eqn{e:yc2}, we can see that at this order in $\lambda$, we already have the influence of the charm Yukawa.  
Moreover, noticing the similarity of the leading terms of \Eqs{e:yc2}{e:yc3}, we can isolate it by exploiting
\begin{eqnarray}\label{e:ycrelation}
\left(  V^\dagger \, D_u^2 \,  V \right)_{3,3} \left(  V^\dagger \, D_u^2 \,  V \right)_{2,2}  - \left(  V^\dagger \, D_u^2 \,  V \right)_{2,3}^2 &\approx&  v^2\left[\lambda^6 \, y_c^2 \, y_t^2 + \mathcal{O}\left(\lambda^8\right)\right] \, .
\end{eqnarray}
Similarly to the approach used for \Eqn{e:ytLO}, we can compare this with the corresponding combination of elements in \Eqn{e:matching2} to obtain the leading order contribution to $y_c$:
\begin{subequations}
	\begin{eqnarray}
	\label{e:ycLO}
	&& v^2 \lambda^6 \, y_c^2 \, y_t^2 \approx \left( H_u \right)_{3,3} \left( H_u \right)_{2,2} - \left( H_u \right)^2_{2,3} \,, \\
	\Rightarrow  && y_c^2 \approx \frac{v_3^2}{ v^2} \, \frac{\epsilon_2^2}{\lambda^6} \, \dfrac{y_{22} \, y_{33} -  y_{23}^2}{ y_t^2}  \,.
	\end{eqnarray}
\end{subequations}
Now using \Eqn{e:suppressions} in conjunction with \Eqn{e:eps12} we can write
the charm quark mass as follows:
\begin{eqnarray}
\label{e:mc}
m_c \equiv \frac{1}{\sqrt{2}}\, y^u_2 \, v_2 \,, \quad {\rm with} \quad 
y^u_2 = \sqrt{\dfrac{y_{22} \, y_{33} -  y_{23}^2}{ y_t^2}}  \,,
\end{eqnarray}
which should make obvious the fact that $v_2$ is the dominant source of $m_c$.
Finally, we need to estimate the mass of the up quark. The simplest way to do it
will be to note from \Eqn{e:matching1} that
\begin{eqnarray}
\label{e:det}
	{\rm Det} (H_u) = {\rm Det} (D_u^2) \equiv m_u^2 \, m_c^2 \, m_t^2 \,.
\end{eqnarray}
Using \Eqn{e:matching2} we can easily calculate
\begin{subequations}
	\begin{eqnarray}
	&& {\rm Det}(H_u) = v_3^6 \epsilon_1^2 \epsilon_2^4 f(y_{jk}) \,, \\
	{\rm with,} && f(y_{jk}) = \left[y_{11} \left(y_{22}y_{33}-y_{23}^2\right) -y_{12}^2y_{33} 
	+2 y_{12}y_{13}y_{23} -y_{13}^2y_{22} \right] \,.
	\end{eqnarray}
\end{subequations}
Now we can use \Eqn{e:det} together with Eqs.~(\ref{e:mt}), (\ref{e:mc}) and (\ref{e:eps12}) to
obtain
\begin{eqnarray}
\label{e:mu}
m_u \equiv \frac{1}{\sqrt{2}}\, y^u_1 \, v_1 \,, \quad {\rm with} \quad 
y^u_1 = \dfrac{2\sqrt{2f(y_{jk})}}{y^u_2 y^u_3}  \,.
\end{eqnarray}
Thus collecting Eqs.~(\ref{e:mt}), (\ref{e:mc}) and (\ref{e:mu}) together we can write the 
diagonal mass matrix in the up-quark sector in a form analogous to \Eqn{e:GammaShapes}
\begin{eqnarray}
\label{e:mup}
\dfrac{1}{\sqrt{2}} \mathrm{diag}\left( v_1\, y^u_1 \, , \, \, v_2 \, y^u_2 \, , \, \, v_3 \,  y^u_3 \right)=\mathrm{diag}\left( m_u \, , \, \, m_c \, , \, \, m_t \right) \, . 
\end{eqnarray}

\subsection{Structures of the neutral current matrices}
Since the matrices $N_{1,2}^{u,d}$ defined in \Eqn{e:NuNddef} govern the Yukawa
couplings of the nonstandard scalars, it is important to investigate the detailed
structures of these matrices. As we will show, these matrices in this model are fully
determined in terms of the quark masses and mixings and $\tan\beta_{1,2}$~\cite{Botella:2009pq}.

Since the Yukawa matrices in the down-quark sector are diagonal from the very
beginning, the forms of $N_{1,2}^{d}$ are particularly simple.
Plugging the textures of \Eqn{e:GammaShapes} into the definitions of \Eqn{e:NuNddef} and then using the explicit form of $O_\beta$ in \Eqn{e:Ob}, we can easily
find:
\begin{subequations}
	\label{e:Nd}
	\begin{eqnarray}
	N_1^d&=& \mathrm{diag}\left( \frac{m_d (O_\beta)_{21}}{v_1}, \frac{m_s (O_\beta)_{22}}{v_2}, \frac{m_b (O_\beta)_{23}}{v_3}\right) 
	= 
	\mathrm{diag}\left(-\dfrac{m_d}{v}\dfrac{t_{\beta_1} t_{\beta_2} }{s_{\beta_2}} ,\, \dfrac{m_s}{v}\dfrac{t_{\beta_2}}{t_{\beta_1} s_{\beta_2}}, \, 0\right)  , \\
	N_2^d &=&  \mathrm{diag}\left( \frac{m_d (O_\beta)_{31}}{v_1}, \frac{m_s (O_\beta)_{32}}{v_2}, \frac{m_b (O_\beta)_{33}}{v_3}\right) 
	= \mathrm{diag} \left( - \frac{m_d}{v} t_{\beta_2} ,\, -\frac{m_s}{v} t_{\beta_2} ,\, \frac{m_b}{v} \frac{1}{t_{\beta_2}} \right) .
	\end{eqnarray}
\end{subequations}
%
Clearly, the down-quark sector does not exhibit any tree-level FCNCs mediated by the nonstandard scalars. Furthermore, while the nonstandard neutral Higgs couplings are flavor-diagonal in the down-quark sector, they are not flavor universal.
To elucidate this feature further, we can define a parameter~\cite{Das:2018qjb}
\begin{eqnarray}
	{\cal F}_k^q = \frac{(N_k^d)_{qq}}{m_q} \,,
\end{eqnarray}
with $k=1,2$ and $q=d,s,b$. Unlike in the NFC versions of 3HDMs, the value of the quantity ${\cal F}_k^q$ will crucially depend on the flavor, $q$,
as can be explicitly checked using \Eqn{e:Nd}.
 This characteristic can be utilized, for instance, to selectively enhance the coupling strengths of the second-generation fermions relative to those with the third generation. As already explained, these flavor nonuniversal couplings distinguish the current model from the corresponding NFC versions of 3HDMs.

Let us now proceed to find the neutral current matrices in the up-quark sector.
To do that, we first explicitly define the following projection matrices:
\begin{eqnarray}
	\label{e:P}
	P_1 = {\rm diag}\left(1,\, 0,\, 0 \right), \quad
	P_2 = {\rm diag}\left(0,\, 1,\, 0 \right), \quad
	P_3 = {\rm diag}\left(0,\, 0,\, 1 \right).
\end{eqnarray}
From \Eqs{e:deltexture}{e:MnMp} we can then write
(no summation over $k$)
\begin{eqnarray}
	\Delta_k = \frac{\sqrt{2}}{v_k}P_k\cdot M_p \,,  \qquad {\rm for}~ k =1,2,3 \,. 
\end{eqnarray}
Substituting this into the definition of $N_1^u$ in \Eqn{e:Nu1d1def}
 and then using \Eqn{e:DdDu}, we get
\begin{eqnarray}
	N_1^u = {\cu}_u \left(\sum_{k=1}^{3}\frac{(O_\beta)_{2k}}{v_k}P_k \, M_p  \right) {\cv}_u^\dagger \, = \sum_{k=1}^{3}\frac{(O_\beta)_{2k}}{v_k} \, {\cu}_u \, P_k \, M_p \, {\cv}_u^\dagger = \sum_{k=1}^{3}\frac{(O_\beta)_{2k}}{v_k} \, {\cu}_u \, P_k \, {\cu}_u^\dagger \, D_u \, .
\end{eqnarray}
Now we can use \Eqn{e:VCKMdef} to write (no summation over $k$),
\begin{eqnarray}
	({\cu}_u P_k \, {\cu}_u^\dagger D_u)_{ab} \equiv (V  P_k V^\dagger D_u)_{ab}=(V)_{ak}(V)^*_{bk}(D_u)_{bb} \, .
\end{eqnarray}
Using this we can express the elements of $N_1^u$ in a simple form as
follows: 
\begin{subequations}
	\label{e:Nu}
\begin{eqnarray}
\label{e:Nu1}
(N_1^u)_{ab}= \sum_{k=1}^{3}\frac{(O_\beta)_{2k}}{v_k} (V)_{ak}(V)^*_{bk}(D_u)_{bb} = \left(\frac{-t_{\beta_1} t_{\beta_2}}{s_{\beta_2} } (V)_{a1}(V)^*_{b1} + \frac{t_{\beta_2}}{t_{\beta_1} s_{\beta_2}} (V)_{a2}(V)^*_{b2} \right) \frac{(D_u)_{bb}}{v} \,,
\end{eqnarray}
where the explicit form of $D_u$ has been given in \Eqn{e:DdDu}.
Following similar steps, the elements of $N_2^u$ will be given by\footnote{
It may be noted that, in the limit $V=1$, the expressions of $N_{1,2}^u$ in \Eqn{e:Nu} take the same form
as those of $N_{1,2}^d$ in \Eqn{e:Nd}.
}
\begin{eqnarray}
(N_2^u)_{ab} &=& \sum_{k=1}^{3}\frac{(O_\beta)_{3k}}{v_k} (V)_{ak}(V)^*_{bk}(D_u)_{bb}  \nonumber \\ 
&=& \left(-t_{\beta_2} (V)_{a1}(V)^*_{b1} - t_{\beta_2} (V)_{a2}(V)^*_{b2} + \frac{1}{t_{\beta_2}}(V)_{a3}(V)^*_{b3}  \right) \frac{(D_u)_{bb}}{v} \, .
\label{e:Nu2}
\end{eqnarray}
\end{subequations}
For the sake of completeness, we also provide the explicit expressions for the
neutral current matrices in the charged lepton sector, which will resemble
$N_{1,2}^d$ in \Eqn{e:Nd} and are given by
\begin{subequations}
	\label{e:Nl}
	\begin{eqnarray}
	N_1^\ell&=& \mathrm{diag}\left( \frac{m_e (O_\beta)_{21}}{v_1}, \frac{m_\mu (O_\beta)_{22}}{v_2}, \frac{m_\tau (O_\beta)_{23}}{v_3},\right) 
	= 
	\mathrm{diag}\left(-\dfrac{m_e}{v}\dfrac{t_{\beta_1} t_{\beta_2} }{s_{\beta_2}} ,\, \dfrac{m_\mu}{v}\dfrac{t_{\beta_2}}{t_{\beta_1} s_{\beta_2}}, \, 0\right)  , \\
	N_2^\ell &=&  \mathrm{diag}\left( \frac{m_e (O_\beta)_{31}}{v_1}, \frac{m_\mu (O_\beta)_{32}}{v_2}, \frac{m_\tau (O_\beta)_{33}}{v_3},\right) 
	= \mathrm{diag} \left( - \frac{m_e}{v} t_{\beta_2} ,\, -\frac{m_\mu}{v} t_{\beta_2} ,\, \frac{m_\tau}{v} \frac{1}{t_{\beta_2}} \right) .
	\end{eqnarray}
\end{subequations}

\begin{figure}[H]
	\centering
	\includegraphics[width=0.4\textwidth]{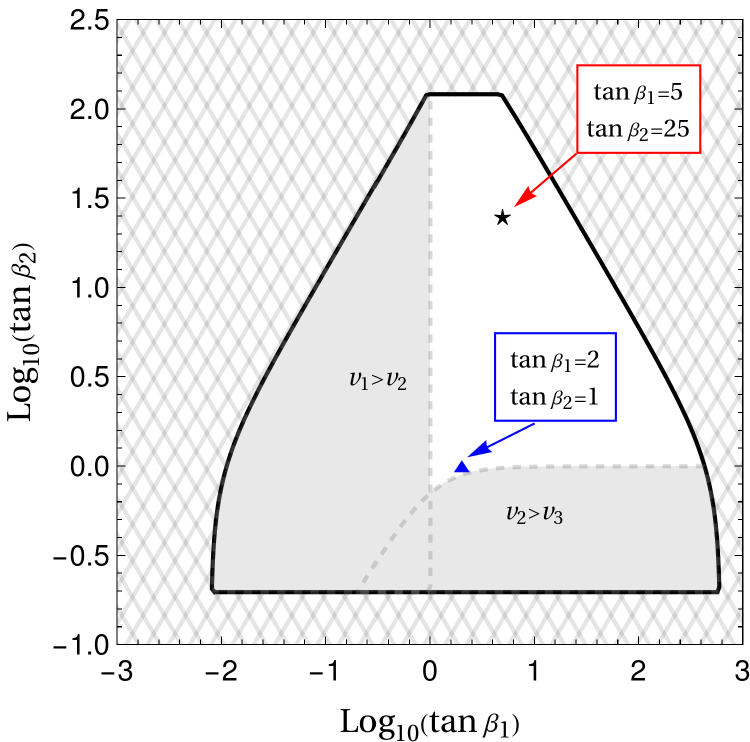}
	\caption{Allowed regions from the perturbativity constraints on the Yukawa couplings. The crossed-out area outside the black boundary
		denotes the combined region forbidden by the constraint $|(N_{1,2}^{u,d,\ell})_{ab}|\le\sqrt{4\pi}$.
		The shaded areas inside the boundary on the left and on the bottom right denote the regions in which the VEV hierarchies invert: $v_1 > v_2$ (left) and $v_2 >v_3$ (bottom right).  
	The benchmark value $\tan\beta_1\approx \order(1/\lambda)\approx 5$ and $\tan\beta_2\approx \order(1/\lambda^2)\approx 25$ is marked with a star, and is inspired from \Eqn{e:vevorders}.
	We also show with a blue triangle the benchmark $\tan\beta_1 = 2$ and $\tan\beta_2=2$, in which the VEV hierarchies are relaxed.} 
	\label{f:pert}
\end{figure}

It should be emphasized that the expressions for
the matrices $N_{1,2}^{u,d,\ell}$ in Eqs.~(\ref{e:Nd}), (\ref{e:Nu}) and (\ref{e:Nl}) are completely general 
and valid for any values of $\tan\beta_{1,2}$. Therefore, although the hierarchy $v_1\ll v_2\ll v_3$ is desirable from
an aesthetic viewpoint, it is not essential for the phenomenological viability of the model.
In fact, in Fig.~\ref{f:pert} we have displayed the region in the $\tan\beta_1$-$\tan\beta_2$ plane
consistent with the perturbativity of the nonstandard Yukawa couplings, $|(N_{1,2}^{u,d,\ell})_{ab}|\le\sqrt{4\pi}$.
As shown in Fig.~\ref{f:pert}, the benchmark sample ($\tan\beta_1= 5$, $\tan\beta_2= 25$), which is
compatible with the hierarchy $v_1\ll v_2\ll v_3$, lies towards the boundary of the perturbative regime,
where the Yukawa couplings of the nonstandard scalars are presumably large and of $\order(1)$. Therefore,
for such values of $\tan\beta_{1,2}$, flavor constraints are expected to impose strong bounds on the nonstandard masses.
As a result, portions of the allowed parameter space in Fig.~\ref{f:pert} will be excluded once flavor constraints are applied. As we will see in the next section, 
the exact extent to which the allowed region in Fig.~\ref{f:pert} shrinks due to the flavor constraints will depend crucially on how light the nonstandard masses that we wish to accommodate are.

\section{Constraints from flavor data}
\label{s:flavor}
The severity of the constraints on the nonstandard masses from the flavor data will depend on
how strongly the nonstandard scalars couple to the SM fermions, especially the quarks.
Since the Yukawa couplings are completely determined if we specify $\tan\beta_{1,2}$, we provide the magnitudes
of the elements of $N_{1,2}^{u,d}$ for two sets of benchmark values of $\tan\beta_{1,2}$.
For the benchmark ($\tan\beta_1= 5$, $\tan\beta_2= 25$) implying $v_1\approx 2$~GeV, $v_2\approx 10$~GeV
and $v_3\approx 246$~GeV we have
\begin{subequations}
\label{e:NuNd1}
\begin{eqnarray}
|N_1^{d}| &\approx& \begin{pmatrix}
2.54 \times 10^{-3} & 0 & 0 \\
0 & 2.03 \times 10^{-3} & 0 \\
0 & 0 & 0
\end{pmatrix}, \\
|N_2^{d}| &\approx& \begin{pmatrix}
5.08 \times 10^{-4} & 0 & 0 \\
0 & 1.02 \times 10^{-2} & 0 \\
0 & 0 & 6.83 \times 10^{-4}
\end{pmatrix},  \\
|N_1^{u}| &\approx& \begin{pmatrix}
9.64 \times 10^{-4} & 1.51 \times 10^{-1} & 7.63 \times 10^{-1} \\
2.32 \times 10^{-4} & 8.37 \times 10^{-3} & 6.98 \times 10^{-2} \\
8.85 \times 10^{-6} & 5.26 \times 10^{-4} & 5.32 \times 10^{-4}
\end{pmatrix}, \\
|N_2^{u}| &\approx& \begin{pmatrix}
2.03 \times 10^{-4} & 2.07 \times 10^{-5} & 6.54 \times 10^{-2} \\
3.18 \times 10^{-8} & 1.32 \times 10^{-1} & 7.33 \times 10^{-1} \\
7.59 \times 10^{-7} & 5.53 \times 10^{-3} & 2.92 \times 10^{-3}
\end{pmatrix}.
\end{eqnarray}
\end{subequations}
For a second benchmark ($\tan\beta_1= 2$, $\tan\beta_2= 1$) implying $v_1\approx 78$~GeV, $v_2 \approx 156$~GeV
and $v_3\approx 174$~GeV, which follows a mild hierarchy $v_1< v_2< v_3$, we will have
\begin{subequations}
	\label{e:NuNd2}
	\begin{eqnarray}
	|N_1^{d}| &\approx & \begin{pmatrix}
	5.74 \times 10^{-5} & 0 & 0 \\
	0 & 2.87 \times 10^{-4} & 0 \\
	0 & 0 & 0
	\end{pmatrix}, \\
	|N_2^{d}| &\approx& \begin{pmatrix}
	2.03 \times 10^{-5} & 0 & 0 \\
	0 & 4.06 \times 10^{-4} & 0 \\
	0 & 0 & 1.71 \times 10^{-2}
	\end{pmatrix}, \\
	|N_1^{u}| &\approx& \begin{pmatrix}
	2.15 \times 10^{-5} & 4.09 \times 10^{-3} & 2.09 \times 10^{-2} \\
	6.30 \times 10^{-6} & 2.76 \times 10^{-3} & 1.64 \times 10^{-2} \\
	2.42 \times 10^{-7} & 1.24 \times 10^{-4} & 6.91 \times 10^{-4}
	\end{pmatrix}, \\
	|N_2^{u}| &\approx& \begin{pmatrix}
	8.13 \times 10^{-6} & 1.65 \times 10^{-6} & 5.23 \times 10^{-3} \\
	2.54 \times 10^{-9} & 5.27 \times 10^{-3} & 5.86 \times 10^{-2} \\
	6.06 \times 10^{-8} & 4.42 \times 10^{-4} & 6.98 \times 10^{-1}
	\end{pmatrix}. 
	\end{eqnarray}
\end{subequations}
As we can explicitly see from \Eqs{e:NuNd1}{e:NuNd2}, the nonstandard Yukawa couplings are much stronger
in the case when $v_1\ll v_2\ll v_3$. Therefore, if we wish to safeguard relatively light nonstandard scalars
against the flavor constraints, we will need to relax the strong VEV hierarchy.

We will analyze the flavor constraints in the alignment limit defined in \Eqn{e:aligncond}.
In addition, to ensure automatic compatibility with the theoretical constraints from
unitarity\cite{Das:2014fea,Bento:2022vsb} and boundedness from below\cite{Faro:2019vcd}, we impose the following 
relations:\footnote{Under these convenient assumptions the scalar potential in \Eqn{e:potential} will
reduce to the maximally symmetric form~\cite{Chakraborti:2021bpy,Darvishi:2021txa}.}
\begin{eqnarray}
\label{e:truncated}
	m_{H_1} = m_{A_1} = m_{C_1} \equiv M_1 \,, \quad
	m_{H_2} = m_{A_2} = m_{C_2} \equiv M_2 \,, \quad
	\alpha_3 = \gamma_1 = \gamma_2 \equiv \gamma \,.
\end{eqnarray}
Because of the above relations, the custodial symmetry will be embodied by the scalar
potential and therefore the constraints arising from the electroweak $T$-parameter
will also be automatically satisfied\cite{Das:2022gbm}.
\begin{figure}[htbp!]
	\centering
	\begin{subfigure}[b]{\textwidth}
		\centering
		\includegraphics[width=.15\textwidth]{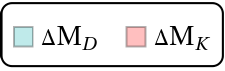}
	\end{subfigure}	
		\includegraphics[width=.9\textwidth]{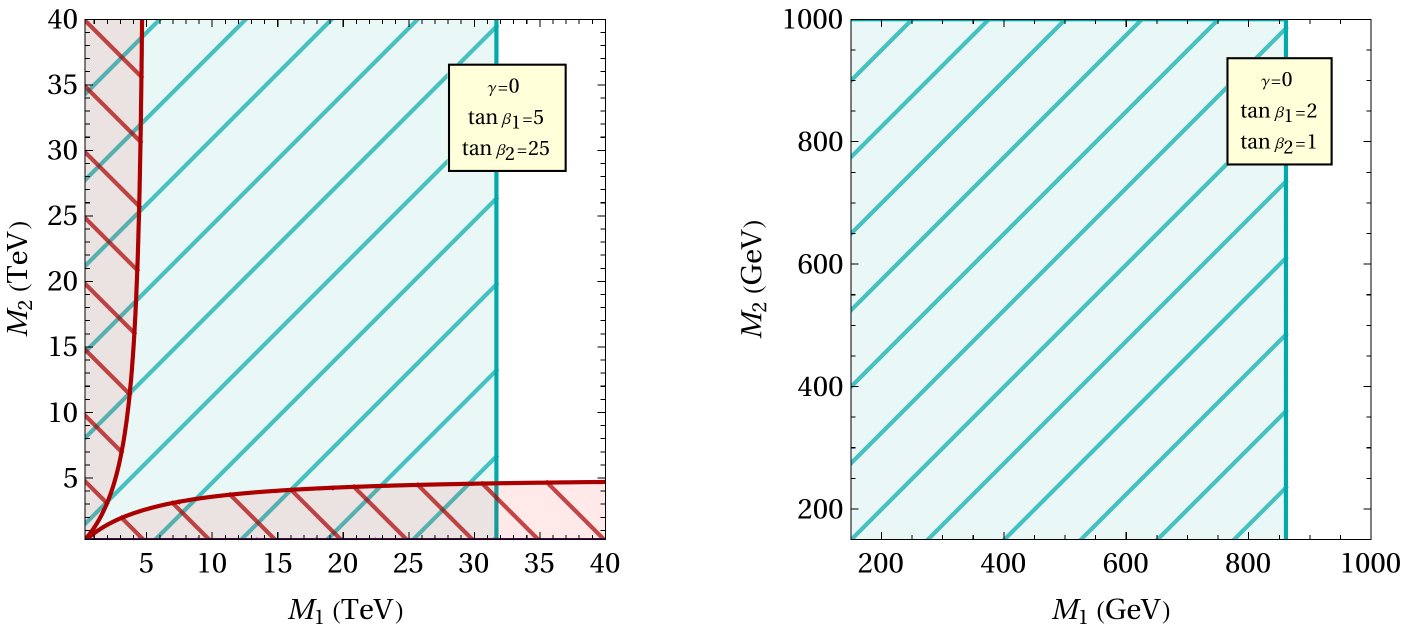}
	\hfill
	\caption{Constraints from flavor data for the case of $\gamma=0$ with $\tan\beta_1 = 5$, $\tan\beta_2 = 25$ (\textbf{left}), and $\tan\beta_1 = 2$, $\tan\beta_2 = 1$ (\textbf{right}). The shaded region is forbidden by $\Delta M_D$ (cyan) and $\Delta M_K$ (red). No additional constraints arise from $\Delta M_{B_{s,d}}$ and $b\to s \gamma$ for the displayed range of parameters.
		Please note that the ranges are different in the axes of the two plots.}
	\label{f:mass-mass}
\end{figure}

We analyze the constraints arising from neutral meson oscillations, namely $\Delta M_K$, $\Delta M_{B_s}$, $\Delta M_{B_d}$, and $\Delta M_{D}$, as well as from $b \to s \gamma$. 
We will focus on the leading order New Physics contributions to these processes and use the relevant expressions and methodologies from Refs.~\cite{Crivellin:2013wna,Botella:2015hoa,Bhattacharyya:2014nja,Nebot:2015wsa}.
First we will address the constraints stemming from $D_0$-$\bar{D_0}$ mixing
which occurs at the tree-level due to the off-diagonal elements $(N_{1,2}^u)_{12}$ and $(N_{1,2}^u)_{21}$.
From \Eqn{e:NuNd1} we can see that $(N_{1}^u)_{12}$ is particularly large for $\tan\beta_{1,2}\gg 1$, a fact that can be easily understood from the explicit expression in \Eqn{e:Nu1}.  
Consequently, it is preferable to selectively decouple one tier of nonstandard scalars, and arrange $\gamma$ such that $N^u_1$ couples predominantly to these
decoupled scalars.  
Based on this strategy, we present benchmark scenarios that largely follow this assumption, with $M_1 \gg M_2$ along with $\gamma \approx 0$. 

%
%

To showcase the flavor constraints for $v_1\ll v_2\ll v_3$,
we begin by setting the benchmark at $\tan\beta_1 \approx 5$ and $\tan\beta_2 \approx 25$,  and investigate the bounds on $M_1$ and $M_2$.  
As already discussed, we set $\gamma=0$ to explicitly see that the $D_0$-$\bar{D_0}$ mixing still allows the second tier of scalars to remain light while $M_1$
is required to be much heavier than the TeV scale.  
Interestingly, no additional constraints arise from the $B_s$ and $B_d$ systems. However,  $K^0 - \bar{K^0}$ prevents $M_2$ from being arbitrarily light, as shown in the left panel of Fig.~\ref{f:mass-mass}.  
For the benchmark $\tan\beta_1 \approx 5$ and $\tan\beta_2 \approx 25$ 
the strength of the FCNC interactions become substantial (see \Eqn{e:NuNd1}), pushing the nonstandard mass scales to large values in order to satisfy the flavor constraints.  
Specifically, from the left panel of Fig.~\ref{f:mass-mass} we observe that even after selectively decoupling $M_1$, $M_2$ needs to be heavier than $\order(5~{\rm TeV})$ to avoid conflicts with $\Delta M_K$. 
On the other hand, if we work with a milder VEV hierarchy, $v_1< v_2< v_3$,
reflected by the benchmark $\tan\beta_1 \approx 2$ and $\tan\beta_2 \approx 1$,
then nonstandard masses below the TeV scale can be easily accommodated as can
be seen from the right panel of Fig.~\ref{f:mass-mass}.

\begin{figure}
	\centering
	\begin{subfigure}[b]{\textwidth}
		\centering
		\includegraphics[width=.75\textwidth]{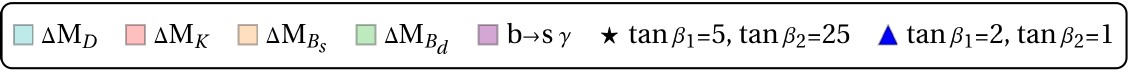}
	\end{subfigure}
	\begin{subfigure}[b]{0.49\textwidth}
		\centering
		\includegraphics[width=1\textwidth]{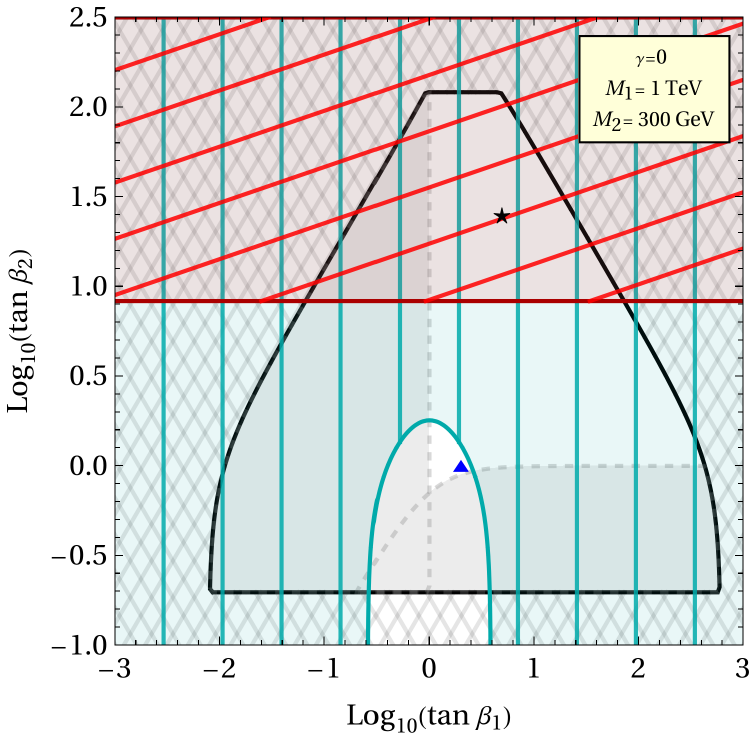}
		\caption[]%
		{{\small $M_1 = 1$ TeV, $M_2 = 300$ GeV, and $\gamma=0$.}}    
		\label{f:bench1}
	\end{subfigure}
	\hfill
	\begin{subfigure}[b]{0.49\textwidth}  
		\centering 
		\includegraphics[width=1\textwidth]{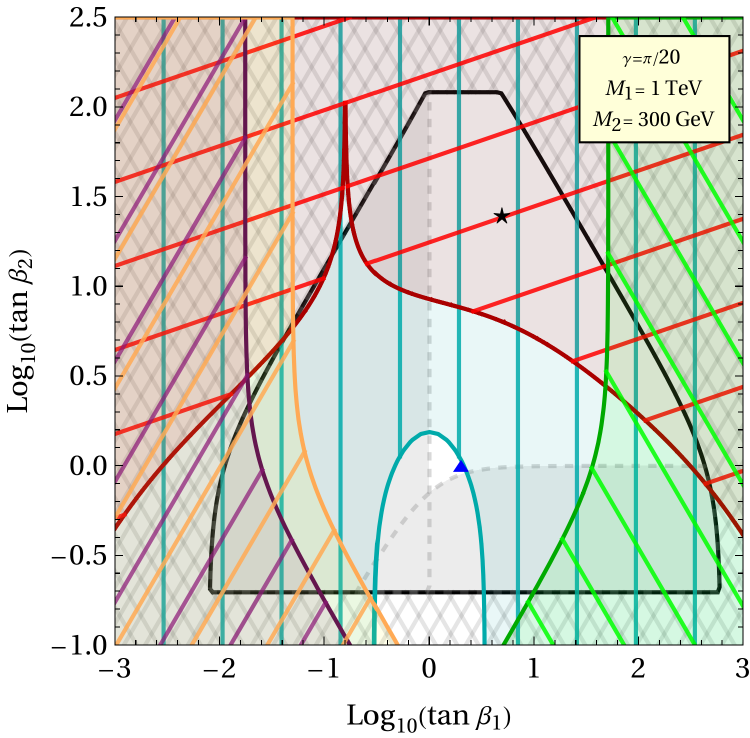}
		\caption[]%
		{{\small  $M_1 = 1$ TeV, $M_2 = 300$ GeV, and $\gamma=\pi/20$.}}    
		\label{f:bench2}
	\end{subfigure}
	\vskip\baselineskip
	\begin{subfigure}[b]{0.49\textwidth}   
		\centering 
		\includegraphics[width=1\textwidth]{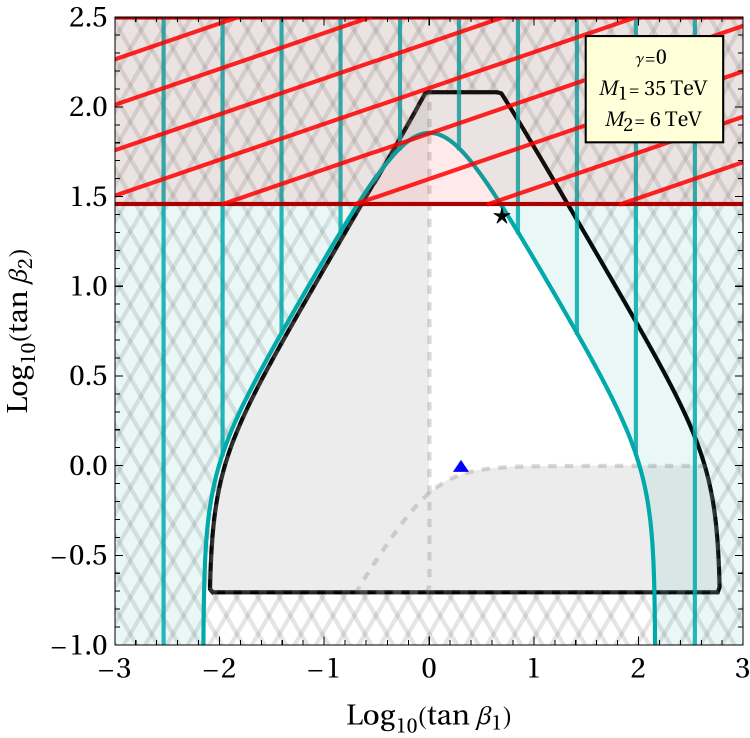}
		\caption[]%
		{{\small  $M_1 = 35$ TeV, $M_2 = 6$ TeV, and $\gamma=0$.}}    
		\label{f:bench3}
	\end{subfigure}
	\hfill
	\begin{subfigure}[b]{0.49\textwidth}   
		\centering 
		\includegraphics[width=1\textwidth]{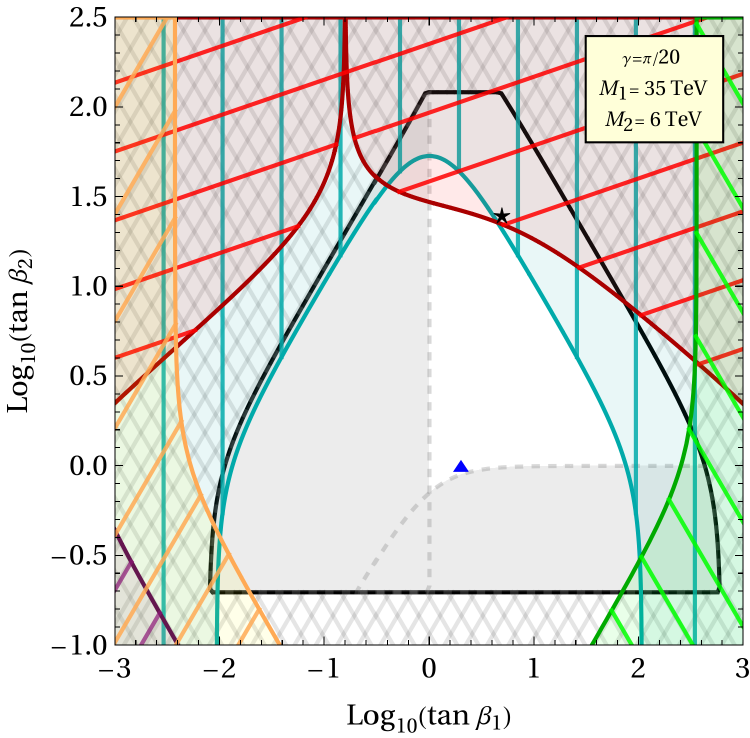}
		\caption[]%
		{{\small  $M_1 = 35$ TeV, $M_2 = 6$ TeV, and $\gamma=\pi/20$.}}    
		\label{f:bench4}
	\end{subfigure}
	\caption[]
	{\small Flavor data constraints on the $\tan\beta_1$-$\tan\beta_2$ plane for different benchmarks of masses and mixing.  
		The regions shaded and crossed out are excluded due to $\Delta M_D$ (cyan),  $\Delta M_K$ (red),  $\Delta M_{B_s}$ (orange),  $\Delta M_{B_d}$ (green),  and $b \to s \gamma$ (purple).  
		The perturbative region shown in Fig.~\ref{f:pert} is overlaid, with the region outside the black boundary being non-perturbative.  The light-gray regions are actually allowed from perturbativity, but correspond to $v_1>v_2$ (left) and $v_2>v_3$ (bottom right).  The star pinpoints the value $\tan\beta_1 = 5$ and $\tan\beta_2 = 25$, whereas the blue triangle marks the $\tan\beta_1 = 2$, $\tan\beta_2 = 1$ point.} 
	\label{f:benchmarks}
\end{figure}

We will now cast these bounds in the $\tan\beta_1$-$\tan\beta_2$ plane
to illustrate 
the extent to which the flavor constraints exclude regions of the allowed parameter space shown in Fig.~\ref{f:pert}.
As previously noted, it is not mandatory
to have $M_1, M_2 \gg v$ to satisfy the flavor constraints.
By relaxing the hierarchy between the VEVs, the model can still accommodate nonstandard scalars below the TeV scale, as showcased in Figs.~\ref{f:bench1} and \ref{f:bench2}. 
In these figures, we observe that even for masses as light as $M_1=1$ TeV and $M_2=300$ GeV, the flavor constraints can be successfully satisfied in the unshaded region characterized
by $\tan\beta_{1,2}\approx\order(1)$ which corresponds to a relatively mild VEV hierarchy
$v_1<v_2<v_3$.  
The differences between Figs.~\ref{f:bench1} and \ref{f:bench2} highlight the fact that even a small mixing can significantly affect the allowed parameter space.  
However, if we insist on a stronger VEV hierarchy, $v_1\ll v_2\ll v_3$, then
the flavor constraints will force the nonstandard masses to be much heavier
than the electroweak scale. This is clearly illustrated in Figs.~\ref{f:bench3} and \ref{f:bench4} where it is evident that the nonstandard masses must lie
well above the TeV scale so that the benchmark $\tan\beta_1 = 5$, $\tan\beta_2 = 25$
may be accommodated within the allowed region. Even in this case, it is important to choose $\gamma\approx 0$ to minimize the impact of the flavor constraints.

\section{Summary}
\label{s:summary}

BGL-3HDMs feature intriguing Yukawa textures, where the $k$-th generation of the left-handed fermion doublets coupling exclusively to the $k$-th scalar doublet.  
This behavior, together with the observed hierarchical quark masses and mixings, allows the primary sources of mass for different generations of fermions to be disentangled.  
The distinction between generations, made possible by the flavor symmetry, becomes particularly evident by the $U(1)_1\times U(1)_2$ realization of the model.  
Notably, this disentanglement of mass sources does not introduce additional parameters stemming purely from the Yukawa sector.  
In the version of the model considered here, the Yukawa matrices in the down-quark and charged lepton sectors are manifestly diagonal, simplifying the phenomenology.
This lack of off-diagonal elements eliminates tree-level FCNCs in these sectors, 
and the Yukawa couplings can be directly related to the physical masses
leading to a more straightforward interpretation.  
All the tree-level FCNC interactions mediated by neutral nonstandard Higgses, in the current model, are confined to the up-quark sector only.  
Moreover, $\tan\beta_1$ and $\tan\beta_2$ are the only new BSM parameters that appear in the expressions for $N_{1,2}^{u, d, \ell}$. 
This particular feature makes these models as predictive as the NFC versions of 3HDM
with a $U(1)_1\times U(1)_2$ symmetry.

Contrary to our initial impressions, the flavor constraints  turned out to
be very strong for the up-type BGL-3HDM considered here. For the desired VEV
hierarchy $v_1\ll v_2\ll v_3$, the constraints from neutral meson mixings
forbid nonstandard masses below $\order(5~{\rm TeV})$. To allow nonstandard
scalars in the sub-TeV regime, we need to relax the aesthetic appeal of maintaining the hierarchy $v_1\ll v_2\ll v_3$, {\it i.e.}, $\tan\beta_{1,2}\gg 1$. This is
a direct consequence of the fact that the strength of the tree-level FCNC
interactions in the up-quark sector are proportional to the up-type masses.
In retrospect, we believe the down-type BGL-3HDM has a relatively better chance of accommodating the flavor constraints. This is because, in the down-type BGL-3HDM case, the tree-level FCNC couplings in the down-quark sector are proportional to the down-type quark masses.

To summarize, despite the strong constraints from the flavor observables,
BGL-3HDMs are not without aesthetic appeal.  
The fact that the bounds on the nonstandard scalar masses are pushed upwards if we insist on diluting the hierarchies among the Yukawa parameters,
follows directly from the predictive nature of the FCNC matrices.  
The presence of flavor-diagonal yet nonuniversal couplings in certain fermionic sectors stands out as a key feature of this scenario.  
This offers a promising avenue for future exploration, particularly if signs of flavor nonuniversality emerge in upcoming data.

\section*{Acknowledgments}
 D.D. thanks the Anusandhan National Research Foundation (ANRF), India for financial support through grant no. CRG/2022/000565. 
 AMP acknowledges support from the University Grants Commission, India through the SRF fellowship scheme.

\appendix
\section{Scalar sector details}
\label{app1:ScalarSector}
The scalar potential presented in \Eqn{e:potential} involves fifteen parameters.  
The bilinear coefficients $m_{11}^2$, $m_{22}^2$, and $m_{33}^2$ can be replaced by the three VEVs, $v_1$, $v_2$, and $v_3$ or equivalently by $v$, $\tan\beta_1$, and $\tan\beta_2$.   
 The remaining 12 parameters -- comprising 9 quartic couplings and 3 soft-breaking parameters -- can be mapped onto the seven physical masses (three CP-even scalars, two CP-odd scalars, and two pairs of charged scalars) and five mixing angles (three in the CP-even sector, one in the CP-odd sector, and one in the charged scalar sector). The details of this procedure are provided below.

The minimization conditions used to replace the
bilinear parameters in terms of the VEVs are given by
\begin{subequations}
	\begin{eqnarray}
		m_{11}^2 &=&   -\lambda_1 v_1^2 -\frac{1}{2}\left\{
		(\lambda_4+\lambda_7)v_2^2 +(\lambda_5+\lambda_8)v_3^2
		\right\} + m_{12}^2 \frac{v_2}{v_1} + m_{13}^2 \frac{v_3}{v_1} \,, 
		\\
		m_{22}^2 &=& -\lambda_2 v_2^2 -\frac{1}{2}\left\{
		(\lambda_4+\lambda_7)v_1^2 +(\lambda_6+\lambda_9)v_3^2
		\right\} + m_{12}^2 \frac{v_1}{v_2} + m_{23}^2 \frac{v_3}{v_2}\,, 
		\\
		m_{33}^2 &=& -\lambda_3 v_3^2 -\frac{1}{2}\left\{
		(\lambda_5+\lambda_8)v_1^2 +(\lambda_6+\lambda_9)v_2^2
		\right\} + m_{13}^2 \frac{v_1}{v_3} + m_{23}^2 \frac{v_2}{v_3}\,. 
	\end{eqnarray}
	\label{e:bilinears}
\end{subequations}
Now let us demonstrate the diagonalization of the mass matrices in different sectors 
following the same prescription as in \cite{Das:2019yad}
but in the presence of the soft terms.


\subsection{CP-odd scalar sector} 
The mass terms for the  CP-odd scalars can be extracted from the scalar potential as
\begin{eqnarray}\label{e:cpoddmassmat}
	V_{\mathrm{PS}}^{\rm mass} = \begin{pmatrix}
		z_1 & z_2 & z_3
	\end{pmatrix} \, \frac{{\cal M}_P^2}{2} \, \begin{pmatrix}
		z_1\\  z_2\\ z_3\\
	\end{pmatrix} \,,
\end{eqnarray}
where ${\cal M}_{P}^2$ is the mass matrix in the Lagrangian basis. It can be block diagonalized by the rotation $\mathcal{O}_\beta$, defined in \Eqn{e:Ob}, as follows:
\begin{subequations}
	\begin{eqnarray}\label{e:mp2x2}
		({\cal B}_{P})^2 \equiv {\cal O}_{\beta} \cdot {\cal M}_{P}^2 \cdot {\cal O}_{\beta}^{T} &=& \begin{pmatrix}
			0 & 0 & 0 \\
			0 & {({\cal B}_P^2)}_{22}  & {({\cal B}_P^2)}_{23} \\
			0 & {({\cal B}_P^2)}_{23} & {({\cal B}_P^2)}_{33} \\
		\end{pmatrix} \,.
	\end{eqnarray}
	The elements of the block diagonal matrix are given by,
	\begin{eqnarray}
		{({\cal B}_P^2)}_{22} &=& \frac{ m_{12}^2 \left(v_1^2+v_2^2\right){}^2+v_3 \left( m_{23}^2 v_1^3+ m_{13}^2 v_2^3\right)}{ v_1 v_2 \left(v_1^2+v_2^2\right)} \,, 
		\\
		{({\cal B}_P^2)}_{23}  &=&\frac{v \left( m_{13}^2 v_2 -m_{23}^2 v_1\right)}{ \left(v_1^2+v_2^2\right)} \,,\\ 
		{({\cal B}_P^2)}_{33} &=&  -\frac{v^2 \left( m_{13}^2 v_1+ m_{23}^2 v_2\right)}{ \left(v_1^2+v_2^2\right) v_3} \,.
	\end{eqnarray}\label{e:BP2}
\end{subequations}
Further, the matrix ${{\cal B}_P^2}$ can be diagonalized by an orthogonal rotation $\mathcal{O}_{\gamma_1}$ as follows:
\begin{eqnarray}
	\label{e:BProt}
	{\cal O}_{\gamma_1} \cdot ({\cal B}_{P})^2 \cdot {\cal O}_{\gamma_1}^T &=& 
	{\rm diag}  (0,~ m^2_{A_1},~ m^2_{A_2})
	\,,\label{e:PSrot} 
\end{eqnarray}
where, ${\cal O}_{\gamma_1}$ is defined as 
	\begin{eqnarray}
	{\cal O}_{\gamma_1} = 
	\begin{pmatrix}
	1 & 0 & 0 \\ 
	0 & \cos\gamma_1 & -\sin\gamma_1 \\ 
	0 & \sin\gamma_1 & \cos\gamma_1 \end{pmatrix}  \label{e:Ogamma1} \,.
\end{eqnarray}
This gives us the following relations
\begin{subequations}\label{e:mAtoBP2}
	\begin{eqnarray}
		m^2_{A_1} \cos^2 \gamma_1 +  m^2_{A_2} \sin^2 \gamma_1 &=& {({\cal B}_P^2)}_{22}   \,, \\
		\cos \gamma_1 \sin\gamma_1 (m^2_{A_2} - m^2_{A_1})  &=&	{({\cal B}_P^2)}_{23} \,, \\
		m^2_{A_1} \sin^2 \gamma_1 +  m^2_{A_2} \cos^2 \gamma_1 &=& {({\cal B}_P^2)}_{33}  \,.
	\end{eqnarray}
\end{subequations}
Additionally,we define the following dimensionless quantities to reparameterize the soft breakings:
\begin{subequations}
	\begin{eqnarray}
		s_{12}=\frac{2m_{12}^2}{v_1 v_2} ~ , ~ s_{13}=\frac{2m_{13}^2}{v_1 v_3} ~ , ~ s_{23}=\frac{2m_{23}^2}{v_2 v_3} \, .
	\end{eqnarray}
\end{subequations}
Using Eq.~(\ref{e:BP2}), Eq.~(\ref{e:mAtoBP2}) we can solve for $s_{12}\,, s_{13}$ and $s_{23}$ as
\begin{subequations}\label{e:sij}
	\begin{eqnarray}
			s_{12}&=& \frac{m_{A_1}^2+m_{A_2}^2}{2v^2} \left(\frac{1}{c_{\beta _2}^2}-\frac{s_{\beta _2}^2}{c_{\beta _2}^2}+1\right) \nonumber \\ 
			&& + \frac{m_{A_1}^2-m_{A_2}^2}{2v^2} \left(-\frac{c_{2 \beta _2} c_{\gamma _1}^2}{c_{\beta _2}^2}+\frac{3 c_{\gamma _1}^2}{c_{\beta _2}^2}-\frac{s_{\beta _2}^2 s_{\gamma _1}^2}{c_{\beta _2}^2}-\frac{3 s_{\gamma _1}^2}{c_{\beta _2}^2}-\frac{2 c_{2 \beta _1} s_{\beta _2} s_{2 \gamma _1}}{c_{\beta _1} c_{\beta _2}^2 s_{\beta _1}}+s_{\gamma _1}^2\right)    \, ,\\
			s_{13}&=&\frac{m_{A_1}^2+m_{A_2}^2}{v^2}+\frac{m_{A_1}^2-m_{A_2}^2}{v^2} \left(-c_{\gamma _1}^2-\frac{s_{\beta _1} s_{2 \gamma _1}}{c_{\beta _1} s_{\beta _2}}+ s_{\gamma _1}^2\right) \, ,\\
			s_{23}&=&\frac{m_{A_1}^2+m_{A_2}^2}{v^2}+\frac{m_{A_1}^2-m_{A_2}^2}{v^2} \left(- c_{\gamma _1}^2+\frac{ c_{\beta _1} s_{2 \gamma _1}}{s_{\beta _1} s_{\beta _2}}+ s_{\gamma _1}^2\right) \,.
	\end{eqnarray}
\end{subequations}
%


\subsection{Charged scalar sector} 
Similar to the pseudoscalar sector case, the charged scalar sector mass matrix ${\cal M}_{C}^2$ can be block diagonalized as
\begin{subequations}\label{e:BC2}
	\begin{eqnarray}
		({\cal B}_{C})^2 \equiv {\cal O}_{\beta} \cdot {\cal M}_{C}^2 \cdot {\cal O}_{\beta}^{T} &=& \begin{pmatrix}
			0 & 0 & 0 \\
			0 & {({\cal B}_C^2)}_{22}  & {({\cal B}_C^2)}_{23} \\
			0 & {({\cal B}_C^2)}_{23} & {({\cal B}_C^2)}_{33} \\
		\end{pmatrix} \,.
	\end{eqnarray}
	where,
	\begin{eqnarray}
		{({\cal B}_C^2)}_{22} &=&\frac{1}{2 \left(v_1^2+v_2^2\right)}\left(s_{23} v_3^2 v_1^2+s_{12} \left(v_1^2+v_2^2\right){}^2 +s_{13} v_2^2 v_3^2 \right. \nonumber \\ && \left. -\lambda _7 v_1^4-2 \lambda _7 v_2^2 v_1^2-\lambda _9 v_3^2 v_1^2-\lambda _7 v_2^4-\lambda _8 v_2^2 v_3^2\right)\,, \\
		{({\cal B}_C^2)}_{23}  &=& \frac{v_1 v_2 v_3}{2 \left(v_1^2+v_2^2\right)}v \left(s_{13} -s_{23} -\lambda _8 +\lambda _9 \right) \,, \\
		{({\cal B}_C^2)}_{33} &=& \frac{v^2}{2 \left(v_1^2+v_2^2\right)} \left(+s_{13} v_1^2+s_{23} v_2^2 -\lambda _8 v_1^2-\lambda _9 v_2^2 \right) \,.
	\end{eqnarray}
\end{subequations}

Further, the charged scalar mass matrix can be completely diagonalized  with the use of the rotation matrix ${\cal O}_{\gamma_2}$ as 
\begin{equation}
	{\cal O}_{\gamma_2} \cdot ({\cal B}_{C})^2 \cdot {\cal O}_{\gamma_2}^T = {\rm diag} (0,~ m^2_{C_1},~ m^2_{C_2}) \, ,
\end{equation}
where, ${\cal O}_{\gamma_2}$ is defined as 
\begin{eqnarray}
	{\cal O}_{\gamma_2} = 
	\begin{pmatrix}
		1 & 0 & 0 \\ 
		0 & \cos\gamma_2 & -\sin\gamma_2 \\ 
		0 & \sin\gamma_2 & \cos\gamma_2 \end{pmatrix}  \label{e:Ogamma2} \,.
\end{eqnarray}
Doing so, we get the following relations:
\begin{subequations}\label{e:mctoapbpcp}
	\begin{eqnarray}
		m^2_{C_1} \cos^2 \gamma_2 +  m^2_{C_2} \sin^2 \gamma_2 &=& {({\cal B}_C^2)}_{22}\,, \\
		\cos \gamma_2 \sin\gamma_2 (m^2_{C_2} - m^2_{C_1}) &=&  {({\cal B}_C^2)}_{23}\,, \\
		m^2_{C_1} \sin^2 \gamma_2 +  m^2_{C_2} \cos^2 \gamma_2 &=&  {({\cal B}_C^2)}_{33}\,.
	\end{eqnarray}
\end{subequations}
These equations in conjunction with Eq.~(\ref{e:BC2}) will enable us to solve for $\lambda_{7},\lambda_8$, and $\lambda_9$ as given below:
\begin{subequations} \label{e:lam789}
	\begin{eqnarray}
		\lambda_7&=& s_{12}-\frac{m_{C_1}^2+m_{C_2}^2}{v^2}+\frac{m_{C_1}^2-m_{C_2}^2}{2 v^2} \left(\frac{c_{2 \beta _2} c_{2 \gamma _2}}{c_{\beta _2}^2}-\frac{3 c_{2 \gamma _2}}{c_{\beta _2}^2}+\frac{2 c_{2 \beta _1} s_{\beta _2} s_{2 \gamma _2}}{c_{\beta _1} c_{\beta _2}^2 s_{\beta _1}}\right) \, , \\
		\lambda_8&=& s_{13}+ \frac{m_{C_1}^2}{v^2} \left(\frac{s_{\beta _1} s_{2 \gamma _2}}{c_{\beta _1} s_{\beta _2}}-2 s_{\gamma _2}^2\right)- \frac{m_{C_2}^2}{v^2} \left(2 c_{\gamma _2}^2+\frac{s_{\beta _1} s_{2 \gamma _2}}{c_{\beta _1} s_{\beta _2}}\right) \, ,\\
		\lambda_9&=& s_{23}-\frac{m_{C_1}^2}{v^2} \left(\frac{c_{\beta _1} s_{2 \gamma _2}}{s_{\beta _1} s_{\beta _2}}+2 s_{\gamma _2}^2\right)+\frac{m_{C_2}^2}{v^2} \left(\frac{c_{\beta _1} s_{2 \gamma _2}}{s_{\beta _1} s_{\beta _2}}-2 c_{\gamma _2}^2\right) \,. 
	\end{eqnarray}
\end{subequations}
where, the expressions for $s_{12},s_{13 }$ and $s_{23}$ can be replaced using Eq.~(\ref{e:sij}). 


\subsection{CP-even scalar sector}
The mass terms of the CP-even scalars in the lagrangian basis can be extracted from the potential as,
\begin{subequations} 
	\begin{eqnarray}\label{e:neutralmassmat}
		V_{S}^{\rm mass} =\begin{pmatrix}
			h_1 & h_2 & h_3\\
		\end{pmatrix} \frac{{\cal M}_S^2}{2} \begin{pmatrix}
			h_1\\  h_2\\ h_3\\
		\end{pmatrix} \,,
	\end{eqnarray}
	where, ${\cal M}_S^2$ is the mass matrix whose elements are as follows,
	\begin{eqnarray}
		{({\cal M}_S^2)}_{11} &=& \frac{1}{2}\left(s_{12} v_2^2 +s_{13} v_3^2 +4 \lambda _1 v_1^2\right) \,, \\
		{({\cal M}_S^2)}_{12} &=& v_1 v_2 \left(-\frac{1}{2} s_{12} +\lambda _4 +\lambda _7 \right) \,, \\
		{({\cal M}_S^2)}_{13} &=& v_1 v_3 \left(-\frac{1}{2} s_{13} +\lambda _5 +\lambda _8  \right) \,, \\
		{({\cal M}_S^2)}_{22} &=& \frac{1}{2}\left(s_{12} v_1^2 +s_{23} v_3^2 +4 \lambda _2 v_2^2\right)\,, \\ 
		{({\cal M}_S^2)}_{23} &=& v_2 v_3 \left(-\frac{1}{2} s_{23} +\lambda _6 +\lambda _9 \right) \,, \\
		{({\cal M}_S^2)}_{33} &=& \frac{1}{2}\left(s_{13} v_1^2 +s_{23} v_2^2 +4 \lambda _3 v_3^2\right) \,.
	\end{eqnarray}\label{e:mselement}
\end{subequations}
This mass matrix can be diagonalized by the following orthogonal rotation,
\begin{eqnarray}\label{e:msdiag}
	{\cal O}_\alpha \cdot {\cal M}_{S}^2 \cdot {\cal O}_\alpha^T &\equiv& \begin{pmatrix}
		m_h^2 & 0 & 0 \\
		0& m_{H_1}^2 & 0 \\
		0 & 0 & m_{H_2}^2 \\
	\end{pmatrix}  \,,
\end{eqnarray}
where, ${\cal O}_\alpha$ is defined as follows:
\begin{subequations}
	\label{e:Oa}
	\begin{eqnarray}
		{\cal O}_\alpha &=& {\cal R}_3 \cdot  {\cal R}_2\cdot {\cal R}_1 \,,
	\end{eqnarray}
with,\footnote{It should be noted that our definition of $\alpha_3$ differs from
that in Ref.~\cite{Das:2019yad} by a negative sign.}
\begin{equation}
\label{e:R}
{\cal R}_1 = \begin{pmatrix}
c_{\alpha_1} & s_{\alpha_1} & 0 \\
-s_{\alpha_1} & c_{\alpha_1} & 0 \\
0 & 0 & 1 \end{pmatrix}, \quad 
{\cal R}_2 = \begin{pmatrix}
c_{\alpha_2} & 0 &s_{\alpha_2}  \\
0 & 1 & 0 \\
-s_{\alpha_2} & 0 & c_{\alpha_2} 
\end{pmatrix},  \quad
{\cal R}_3 = \begin{pmatrix}
1 & 0 & 0 \\
0 & c_{\alpha_3} &  -s_{\alpha_3}  \\
0 & s_{\alpha_3} & c_{\alpha_3 }
\end{pmatrix}.
\end{equation}
\end{subequations}
Inverting Eq.~(\ref{e:msdiag}),
\begin{eqnarray}\label{e:MS}
	{\cal M}_{S}^2  &\equiv& {\cal O}_\alpha^T  \cdot \begin{pmatrix}
		m_h^2 & 0 & 0 \\
		0& m_{H_1}^2 & 0 \\
		0 & 0 & m_{H_2}^2 \\
	\end{pmatrix} \cdot {\cal O}_\alpha\,,
\end{eqnarray}
which can be solved to obtain the remaining six $\lambda_i$ as follows:
\begin{subequations}
	\label{e:lam1to6}
	\begin{eqnarray}
		\lambda_1 &=& -\frac{s_{12} s_{\beta_1}^2}{4 c_{\beta_1}^2}-\frac{s_{13} s_{\beta_2}^2}{4 c_{\beta_1}^2 c_{\beta_2}^2}+\frac{c_{\alpha_1}^2 c_{\alpha_2}^2 m_h^2}{2 v^2 c_{\beta_1}^2 c_{\beta_2}^2}+\frac{m_{H_1}^2}{v^2} \left(\frac{c_{\alpha_3}^2 s_{\alpha_1}^2}{2 c_{\beta_1}^2 c_{\beta_2}^2}+\frac{c_{\alpha_1}^2 s_{\alpha_2}^2 s_{\alpha_3}^2}{2 c_{\beta_1}^2 c_{\beta_2}^2} -\frac{s_{2 \alpha_1} s_{\alpha_2} s_{2 \alpha_3}}{4 c_{\beta_1}^2 c_{\beta_2}^2}\right) \nonumber \\ 
		&& + \frac{m_{H_2}^2}{v^2} \left(\frac{c_{\alpha_1}^2 c_{\alpha_3}^2 s_{\alpha_2}^2}{2 c_{\beta_1}^2 c_{\beta_2}^2} +\frac{c_{\alpha_1} c_{\alpha_3} s_{\alpha_1} s_{\alpha_3} s_{\alpha_2}}{c_{\beta_1}^2 c_{\beta_2}^2}+\frac{s_{\alpha_1}^2 s_{\alpha_3}^2}{2 c_{\beta_1}^2 c_{\beta_2}^2}\right)\, , \\
		\lambda_2&=& -\frac{s_{12} c_{\beta_1}^2}{4 s_{\beta_1}^2} -\frac{s_{23} s_{\beta_2}^2}{4 c_{\beta_2}^2 s_{\beta_1}^2}+\frac{c_{\alpha_2}^2 m_h^2 s_{\alpha_1}^2}{2 v^2 c_{\beta_2}^2 s_{\beta_1}^2}+\frac{m_{H_1}^2}{v^2} \left(\frac{c_{\alpha_1}^2 c_{\alpha_3}^2}{2 c_{\beta_2}^2 s_{\beta_1}^2} +\frac{c_{\alpha_1} c_{\alpha_3} s_{\alpha_1} s_{\alpha_2} s_{\alpha_3}}{c_{\beta_2}^2 s_{\beta_1}^2}+\frac{s_{\alpha_1}^2 s_{\alpha_2}^2 s_{\alpha_3}^2}{2 c_{\beta_2}^2 s_{\beta_1}^2}\right) \nonumber \\
		&&+\frac{m_{H_2}^2}{v^2} \left(\frac{c_{\alpha_3}^2 s_{\alpha_1}^2 s_{\alpha_2}^2}{2 c_{\beta_2}^2 s_{\beta_1}^2}-\frac{s_{2 \alpha_1} s_{2 \alpha_3} s_{\alpha_2}}{4 c_{\beta_2}^2 s_{\beta_1}^2}+\frac{c_{\alpha_1}^2 s_{\alpha_3}^2}{2 c_{\beta_2}^2 s_{\beta_1}^2}\right) \, ,  \\
		\lambda_3&=& -\frac{s_{13} c_{\beta_1}^2 c_{\beta_2}^2}{4 s_{\beta_2}^2} -\frac{s_{23} c_{\beta_2}^2 s_{\beta_1}^2}{4 s_{\beta_2}^2} +\frac{m_h^2 s_{\alpha_2}^2}{2 v^2 s_{\beta_2}^2}+\frac{m_{H_1}^2}{v^2} \left(\frac{c_{\alpha_2}^2}{4 s_{\beta_2}^2}-\frac{c_{\alpha_2}^2 c_{2 \alpha_3}}{4 s_{\beta_2}^2}\right)\nonumber \\
		&& +\frac{m_{H_2}^2}{v^2} \left(\frac{c_{2 \alpha_3} c_{\alpha_2}^2}{4 s_{\beta_2}^2}+\frac{c_{\alpha_2}^2}{4 s_{\beta_2}^2}\right)\, ,\\
		\lambda_4&=& \frac{s_{12}}{2} -\lambda_7 + \frac{c_{\alpha_2}^2 m_h^2 s_{2 \alpha_1}}{2 v^2 c_{\beta_1} c_{\beta_2}^2 s_{\beta_1}} + \frac{m_{H_1}^2}{v^2} \left(-\frac{c_{\alpha_2}^2 s_{2 \alpha_1}}{4 c_{\beta_1} c_{\beta_2}^2 s_{\beta_1}}+\frac{c_{2 \alpha_1} s_{\alpha_2} s_{2 \alpha_3}}{2 c_{\beta_1} c_{\beta_2}^2 s_{\beta_1}}+\frac{c_{2 \alpha_2} c_{2 \alpha_3} s_{2 \alpha_1}}{8 c_{\beta_1} c_{\beta_2}^2 s_{\beta_1}}-\frac{3 c_{2 \alpha_3} s_{2 \alpha_1}}{8 c_{\beta_1} c_{\beta_2}^2 s_{\beta_1}}\right) \nonumber \\ 
		&& + \frac{m_{H_2}^2}{v^2} \left(-\frac{c_{\alpha_2}^2 s_{2 \alpha_1}}{4 c_{\beta_1} c_{\beta_2}^2 s_{\beta_1}}-\frac{c_{2 \alpha_1} s_{\alpha_2} s_{2 \alpha_3}}{2 c_{\beta_1} c_{\beta_2}^2 s_{\beta_1}}-\frac{c_{2 \alpha_2} c_{2 \alpha_3} s_{2 \alpha_1}}{8 c_{\beta_1} c_{\beta_2}^2 s_{\beta_1}}+\frac{3 c_{2 \alpha_3} s_{2 \alpha_1}}{8 c_{\beta_1} c_{\beta_2}^2 s_{\beta_1}}\right)\, ,\\
		\lambda_5&=& \frac{s_{13}}{2} -\lambda_8 +\frac{c_{\alpha_1} m_h^2 s_{2 \alpha_2}}{2 v^2 c_{\beta_1} c_{\beta_2} s_{\beta_2}} + \frac{m_{H_1}^2}{v^2} \left(-\frac{c_{\alpha_1} s_{2 \alpha_2}}{4 c_{\beta_1} c_{\beta_2} s_{\beta_2}}+\frac{c_{\alpha_1} c_{2 \alpha_3} s_{2 \alpha_2}}{4 c_{\beta_1} c_{\beta_2} s_{\beta_2}}+\frac{c_{\alpha_2} s_{\alpha_1} s_{2 \alpha_3}}{2 c_{\beta_1} c_{\beta_2} s_{\beta_2}}\right) \nonumber \\
		&& + \frac{m_{H_2}^2}{v^2} \left(-\frac{c_{\alpha_1} s_{2 \alpha_2}}{4 c_{\beta_1} c_{\beta_2} s_{\beta_2}}-\frac{c_{\alpha_1} c_{2 \alpha_3} s_{2 \alpha_2}}{4 c_{\beta_1} c_{\beta_2} s_{\beta_2}}-\frac{c_{\alpha_2} s_{\alpha_1} s_{2 \alpha_3}}{2 c_{\beta_1} c_{\beta_2} s_{\beta_2}}\right)\, ,\\
		\lambda_6&=& \frac{s_{23}}{2}-\lambda_9+\frac{m_h^2 s_{\alpha_1} s_{2 \alpha_2}}{v^2 s_{\beta_1} s_{2 \beta_2}}+\frac{m_{H_1}^2}{v^2} \left(\frac{c_{2 \alpha_3} s_{\alpha_1} s_{2 \alpha_2}}{2 s_{\beta_1} s_{2 \beta_2}}-\frac{c_{\alpha_1} c_{\alpha_2} s_{2 \alpha_3}}{2 c_{\beta_2} s_{\beta_1} s_{\beta_2}}-\frac{s_{\alpha_1} s_{2 \alpha_2}}{2 s_{\beta_1} s_{2 \beta_2}}\right) \nonumber \\
		&&+ \frac{m_{H_2}^2}{v^2} \left(-\frac{c_{\alpha_3}^2 s_{\alpha_1} s_{2 \alpha_2}}{2 c_{\beta_2} s_{\beta_1} s_{\beta_2}}+\frac{c_{\alpha_1} c_{\alpha_2} s_{2 \alpha_3}}{2 c_{\beta_2} s_{\beta_1} s_{\beta_2}}\right) \,.
	\end{eqnarray} 
\end{subequations}
Substituting $s_{ij}$ from \Eqn{e:sij} and $\lambda_7,\lambda_8,\lambda_9$ from \Eqn{e:lam789}, we obtain expressions for the remaining $\lambda_i$ in terms of the chosen set of parameters.

\bibliographystyle{JHEP}
\bibliography{h3HDM.bib}
\end{document}